\documentclass{article}
\usepackage{amssymb}
\usepackage{cite,multirow} 
\usepackage[dvips]{graphicx,epsfig}
\usepackage{graphics}
\def\1ad{\mbox{\normalsize $^1$}}
\def\2ad{\mbox{\normalsize $^2$}}
\def\3ad{\mbox{\normalsize $^3$}}
\def\4ad{\mbox{\normalsize $^4$}}
\def\5ad{\mbox{\normalsize $^5$}}
\def\6ad{\mbox{\normalsize $^6$}}
\def\7ad{\mbox{\normalsize $^7$}}
\def\8ad{\mbox{\normalsize $^8$}}

\makeatletter
\@addtoreset{equation}{section}
\makeatother
\renewcommand{\theequation}{\thesection.\arabic{equation}}
 
\def\beq{\begin{equation}}                     %  
\def\eeq{\end{equation}}                       % 
\def\bea{\begin{eqnarray}}                     %         % 
\def\eea{\end{eqnarray}}                       %       %  
\def\nn{\nonumber}

\def\0 {\nonumber} 
\def\del{\partial}

\def\A{{\cal A}}
\def\B{{\cal B}}
\def\a{\alpha}
\def\b{\beta}
\def \we { \wedge}
\def \ee {\epsilon}
\def \te {\tilde \epsilon}

\def\minicent#1#2{
  \begin{minipage}{#1 cm}
    \begin{center}
     #2 
    \end{center}
  \end{minipage}
}

\begin{document}

\setcounter{page}{0}
\begin{titlepage}
\titlepage
\rightline{hep-th/0412187}
\rightline{Bicocca-FT-04-18}
\rightline{CPHT-RR 070.1204}
\rightline{LPTENS-04/52}
\vskip 3cm
\centerline{{ \bf \Large The Baryonic Branch of Klebanov-Strassler Solution:}}
\vskip .5cm
\centerline{{ \bf \Large a Supersymmetric Family of SU(3) Structure Backgrounds }}
\vskip 1cm
\centerline{Agostino Butti$^a$, Mariana Gra{\~n}a$^{b,c}$, Ruben
Minasian$^b$, Michela Petrini$^{b,d}$
and Alberto Zaffaroni$^a$}
\vskip 1truecm
\begin{center}
\em 
$^a$ Dipartimento di Fisica, Universit\`{a} di Milano-Bicocca \\ 
P.zza della Scienza, 3; I-20126 Milano, Italy\\
\vskip .4cm
$^b$Centre de Physique Th{\'e}orique, Ecole
Polytechnique
%\footnote{Unit{\'e} mixte du CNRS et de l'EP, UMR 7644}
\\91128 Palaiseau Cedex, France\\
\vskip .4cm
$^c$Laboratoire de Physique Th{\'e}orique, Ecole Normale
Sup{\'e}rieure\\
24, Rue Lhomond 75231 Paris Cedex 05, France\\
\vskip .4cm
$^d$Laboratoire de Math{\'e}matiques et Physique Th{\'e}orique,
 Universit{\'e} Fran{\c c}ois Rabelais\\ 
Parc de Grandmont 37200, Tours, France

\vskip 2cm
\end{center}
\begin{abstract}

\vskip1cm
% Text of abstract in English
We exhibit a one-parameter family of regular supersymmetric solutions of
type IIB theory that describes the baryonic branch of the Klebanov-Strassler 
(KS) theory. The solution is obtained by applying the
supersymmetry conditions for SU(3)-structure manifolds to an interpolating
ansatz proposed by Papadopoulos and Tseytlin. Other than at the KS point, the
family does not have a conformally-Ricci-flat metric, neither it has
self-dual three-form flux. By varying also the string coupling, 
our solution smoothly interpolates between  Klebanov-Strassler and
Maldacena-Nu{\~n}ez (MN). The asymptotic IR and UV are that
of KS throughout the interpolating flow, except for the extremal value of the
parameter where the UV solution drastically changes to MN.

\end{abstract}
\vskip 0.5\baselineskip

\vfill
\begin{flushleft}
%\vspace{.5cm}
\end{flushleft}
\end{titlepage}
\large
\section{Introduction}
 
Supergravity solutions with fluxes are a rich subject that ranges from the
physics of string compactifications and vacua to the AdS/CFT correspondence.
The geometry of such solutions has recently received much attention. Our
improved understanding of this subject reveals that many interesting features
are still to be uncovered and that, most likely, new relevant solutions
are still to be found. In this paper, using the supersymmetry
conditions recently derived in \cite{GMPT}, we study a class of solutions
relevant for the AdS/CFT correspondence. Up to now only two regular
supergravity backgrounds are known that correspond to dual confining 
${\cal N}=1$
SYM theories, the Klebanov-Strassler (KS \cite{KS}) and the Maldacena-Nu\~nez
 (MN \cite{MN}) solutions. It was suggested in \cite{aharony,GHK}
that the KS solution should belong to a one parameter family of supersymmetric
backgrounds describing the baryonic branch of the dual gauge theory. This
expectation was strengthen in  \cite{GHK} (Gubser-Herzog-Klebanov, GHK) 
by a linearized analysis that
exhibits a massless Goldstone boson and a first order deformed solution
of the equations of motion. In this paper we will show that there exists 
indeed a one parameter family of regular supersymmetric deformations of the 
KS solution. The supergravity solution also depends on a second parameter, 
since we are free to add an arbitrary constant to the dilaton. 
The two parameters can be interpreted in the dual gauge theory as
a baryonic VEV and a gauge coupling constant. 
%By fixing the asymptotic value of the dilaton, 
%we obtain the supergravity description of the baryonic branch 
%\footnote{We thank Igor Klebanov
%and Anatoly Dymarsky for a useful discussion about this issue.}.
By varying both parameters, we find a flow whose endpoint is 
the MN solution. The result thus makes justice of the many attempts of 
finding a supergravity solution interpolating between KS and MN 
\cite{GHK,klebstring,FG}.

Supersymmetry in compactifications without fluxes requires the
internal manifold to be a Calabi-Yau. The presence of fluxes back-reacts
on the geometry transforming the manifold into a  generalized Calabi-Yau \cite{GMPT}.
In particular, since supersymmetry requires the existence of at least a
nonvanishing spinor, the internal manifold must have an $SU(3)$ structure.
A set of general conditions that preserve supersymmetry is known for
type II theories. In type IIB theories, the internal manifold must be complex
\cite{GMPT, Frey, DAllagata}.
In addition to this, supersymmetry implies a set of linear
conditions on the $SU(3)$ torsions and the fluxes \cite{GMPT}.
We will apply these supersymmetric constraints 
to an ansatz for the  metric and fluxes proposed by Papadopoulos and Tseytlin (PT \cite{PT}), which contains
both the KS and MN solutions. We will find that, quite amazingly, the
apparently overdetermined set of equations admits a class of solutions.
All unknown functions in the ansatz are explicitly determined
in terms of two quantities for which we provide a system of 
coupled first-order differential equations. The MN and KS backgrounds 
and the linearized deformation of KS found in \cite{GHK} 
are particular solutions of these equations. More generally, by
a numerical and power series analysis, we can show
that there is a one parameter family of regular solutions that deforms
KS, while preserving an $SU(2) \times SU(2)$ symmetry.
We can extend the solution symmetrically with respect to KS, 
but the two branches are related by a $Z_2$ symmetry.

From the field theory perspective, the KS solution is known to represent
an $SU(N+M)\times SU(N)$ ${\cal N}=1$ gauge theory undergoing 
a cascading series of Seiberg dualities that decreases the 
number of colors and terminates with a
single $SU(M)$ gauge group. In the last step of the cascade, the gauge
theory is $SU(2M)\times SU(M)$ and has a moduli space of vacua that
include a baryonic branch. The moduli space is labeled by a complex 
parameter and has a $Z_2$ symmetry. It was proposed in \cite{aharony,GHK} that
the KS solution should describe the symmetric point on this branch. Our
family of solutions, which preserves $SU(2) \times SU(2)$ and has a $Z_2$
symmetry, is the natural candidate for describing the full
baryonic branch of the gauge theory. An analysis of the UV asymptotic
reveals that the set of operators that are acquiring a VEV is compatible
with this interpretation. Moreover, the analysis of the IR behavior
of the metric shows that the physics of confinement is the same as described
in \cite{KS,MN}. In particular, the ratio of string tensions 
\cite{ds,HSZ} smoothly interpolates between the known cases of KS and MN
\cite{herzog}. 

The existence of a family of regular solutions describing 
the moduli space of a gauge theory is particularly significant. 
All other attempts of deforming a known solution along a moduli space always
gave rise to singular backgrounds. For instance, solutions dual to the Coulomb 
branch of vacua in ${\cal N} =4$ or ${\cal N}=2$ theories contain
singularities corresponding to
distributions of branes for whose resolution stringy effects are usually 
invoked \cite{JPP}. 
The solution presented here is the first example of gauge theory 
which can be described in a controllable way along its moduli space.
Only for large values of the baryonic VEV the supergravity
solution becomes strongly coupled \footnote{We thank Igor Klebanov
and Anatoly Dymarsky for a useful discussion about this issue.}.

Moving along the baryonic branch implies
fixing the boundary conditions for the dilaton.
By varying also the string coupling, our solution smoothly interpolates
between KS and MN. All the backgrounds along this flow have the UV  asymptotic 
behavior of the KS solution, except for the extremal value of the interpolating
parameter, where it suddenly changes, the dilaton blows up and we recover 
the MN solution. Through this paper, we will refer to this solution as 
the ``interpolating flow''. From the field theory point of view, we can reach
the MN point by varying both the baryonic VEV and a coupling constant.  

We have also explicitly checked that the equations of motions are automatically satisfied
by the first order system of susy equations we derived.  We are therefore sure that the
one parameter family of solutions we have found are true solutions of type IIB supergravity.

The paper is organized as follows. In Section 2, we review the supersymmetry
conditions for type IIB solutions with $SU(3)$ structure. In Section 3
we present the PT ansatz and we discuss the choice of complex structure on
the manifold. We give the general solution of the supersymmetry conditions and
we show how the known cases (MN, KS and GHK) fit in it. 
The details of the derivation are reported
in the Appendices. In Section 4 we discuss
the family of regular solutions that interpolates between KS and MN.
Finally, in Section 5, we compare our results with the field theory
expectations.

\section{Type IIB solutions with SU(3) structure} 
\label{section2} 

We are interested in type IIB backgrounds with non-zero RR fluxes and 
where the space-time is a warped 
product of the form ${\mathbf R}^{1,3} \times_w M_6$
\beq 
ds^2= e^{2A} \eta_{\mu\nu} dx^{\mu} dx^\nu + ds_6^2 \, ,
\eeq
where $A$ is a function of the internal coordinates. 

For these backgrounds a set of general conditions on the 
six-dimensional manifold 
and the fluxes to preserve supersymmetry is 
known \cite{GMPT, Frey, DAllagata}.
In particular 
supersymmetry forces the manifold $M_6$ to have a globally defined
invariant spinor. This is possible only for manifolds 
that have reduced structure.
The structure group of a manifold is the group of transformations required to
patch the orthonormal frame bundle. 
A six dimensional Riemannian manifold has automatically 
SO(6) structure. In order to preserve supersymmetry, 
the structure group of $M_6$  
should be a subgroup of SO(6). In general, the smaller the group, 
the bigger the number of supercharges of the effective four dimensional theory.
To preserve minimal supersymmetry (${\cal N}=1$ in four dimensions) 
the structure group should be at most SU(3).
The set of supersymmetry conditions for IIB backgrounds 
with  SU(3)-structure was derived in \cite{GMPT}. 
We refer to this paper for the detailed derivation, here  
we briefly summarize the results.
% the supersymmetry conditions for IIB backgrounds 
%with  SU(3)-structure, following \cite{GMPT}. 
For details about SU(3) structure (or G-structures in general) we refer
to  \cite{Salomon,joyce,London} and
references therein.
%, while for the detailed  derivation of the following see \cite{GMPT}.

A manifold with
SU(3) structure has all the group-theoretic features of a Calabi-Yau, 
namely an invariant spinor $\eta$ and two- and three forms, $J$ and $\Omega$ respectively,
that are constructed as bilinears of the spinor. If the 
manifold had SU(3) {\it holonomy}, not only $J$ and $\Omega$ would be
well defined, but also they would be closed: $dJ=0=d\Omega$. If they are not,
$dJ$ and $d\Omega$ give a good measure of how far the manifold is from
having SU(3) holonomy. The failure of an SU(3) structure to become
SU(3) holonomy is measured by the intrinsic torsion. In terms of
SU(3) representations, the intrinsic torsion  has $(3 \oplus \bar{3}
\oplus 1) \otimes (3 \oplus \bar{3})$ components which are defined as
follows:
\begin{equation}
  \label{eq:djdo}
  \begin{array}{c}\vspace{.3cm}
dJ = -\frac{3}{2}\, {\rm Im}(W_1 \bar{\Omega}) + W_4 \wedge J + W_3 \, ,\\
d\Omega = W_1 J^2 + W_2 \wedge J + W_5^{(\bar 3)} \wedge \Omega\ \, .
  \end{array}
\end{equation}
$W_1$ is a complex zero--form in $1 \oplus 1$, $W_2$ is a complex
primitive two--form, so it lies in $8 \oplus 8$, $W_3$ is a real primitive
$(2,1) \oplus (1,2)$ form in the $6 \oplus \bar{6}$, $W_4$ is a 
real one-form in $3 \oplus \bar{3}$ \footnote{In the following we shall decompose the torsions $W_3$, $W_4$ and $W_5$ in the
 complex basis as $W_3=W_3^{(6)}+W_3^{(\bar{6})}$, 
 $W_4=W_4^{(3)}+W_4^{(\bar{3})}$ and  $W_5=W_5^{(3)}+W_5^{(\bar{3})}$.} and finally $ W_5^{(\bar 3)}$ is a complex $(0,1)$-form (notice that in 
(\ref{eq:djdo}) the $(1,0)$ part drops out), so its degrees of 
freedom are again $3 \oplus \bar{3}$. 

$W_1=W_2=0$ corresponds to an integrable complex structure 
($dJ$ does not have (3,0) or (0,3) pieces, and $d\Omega$ misses the (2,2) 
pieces), which should be the case for a complex manifold.
A conformal Calabi-Yau (a space with a conformally-Ricci flat metric) 
has $W_1=W_2=W_3=0$, and non-zero $W_4$ and $W_5$ obeying $3 W_4 = 2  W_5$.

Similarly all the fluxes in the theory can be decomposed in SU(3) 
representations
\begin{equation}
  \label{fluxesdec}
  \begin{array}{c}\vspace{.3cm}
H = -\frac{3}{2}\, {\rm Im}(H^{(1)} \bar{\Omega}) + H^{(3+\bar{3})} \wedge J + 
H^{(6 +\bar{6})} \, , \\
F_3 = -\frac{3}{2}\, {\rm Im}(F_3^{(1)} \bar{\Omega}) + F_3^{(3 + \bar{3})} \wedge J + 
F_3^{(6 + \bar{6})} \, .
  \end{array}
\end{equation}

%\begin{center}
%\begin{tabular}{|c|c|c|c|c|} \hline
% & $1\oplus 1$ & $3\oplus \bar 3$ & $ \   \ \  \ \ 6\oplus \bar 6   \  \
%\ \ \ $ & $8\oplus8 $ \\ \hline
%Torsion & $1 \, (W_1)$ & $2 \, (W_4, W_5)$ & $1 \, (W_3)$ & $ 1 \, (W_2)$
%\\ \hline
%$H$ & $1$ & $1$ & $1$ & $0$ \\ \hline
% $F_{2n+1}$ & $1 \, (F_3) $ & $\ 3 \, (F_1,F_3,F_5)\ $ & $1 \, (F_3)
%$ & $0$ \\ \hline 
%\end{tabular}\\
%\caption{Table 1: Decomposition of torsion and fluxes in SU(3) 
%structure representations.}
%\end{center}

Using group theory it is then possible to reduce the 
supersymmetry conditions on the variations of the fermions to a 
set of algebraic equations for the different SU(3) components of the torsion 
and the fluxes
 
\begin{minipage}{\linewidth}
\beq
\begin{array}{|c||c|c|c|} \hline
& {\cal W} & \rm{NS}  & {\rm RR \, } \\ \hline\hline
1  \oplus {\bar 1} & W_1 & H^{(1)} &  F_3^{(1)} \\
3  \oplus {\bar 3}  & W_4, W_5   & H^{(3)} & F_1^{(3)}, F_3^{(3)}, F_5^{(3)} \\
6 \oplus {\bar 6}  & W_3  & H^{(6)}  & F_3^{(6)}  \\
8 \oplus 8  & W_2 &  -  &    -\\
\hline \end{array}
\eeq
\begin{center}
Table 1: Decomposition of torsion and fluxes in SU(3) representations.
\end{center}
\end{minipage}

Since torsion is induced by the fluxes in order to have a 
supersymmetric solution non-zero torsions must be compensated 
by non-zero fluxes in the same representation.
%The full set of constraints and
%their analysis can be found in \cite{GMPT}, here we only give the results for 
%IIB compactifications.

Table 1 above shows that in IIB theories there are no fluxes to compensate 
torsion in the  $8\oplus8 $. Thus $W_2=0$ in any IIB solution with 
SU(3) structure. 
%There are 42 components of the former and only 32 of the latter, and in 
%particular there is no flux which contains representation 8. 

Supersymmetry imposes additionally
that all the singlets ($W_1,F_3^{(1)}$ and $H^{(1)}$)
vanish
\beq
\label{cond1}
W_1= F_3^{(1)} = H^{(1)}=0 \, .
\eeq
Therefore we get the first basic fact  about IIB backgrounds with SU(3) 
structure - they necessarily 
involve complex manifolds \cite{GMPT,Frey,DAllagata}: 
%In summary, for torsion and fluxes in the first and last column of Table 1,
%supersymmetry imposes
%\beq \label{cond1+8}
%W_1=
%F_3^{(1)}=H^{(1)}=0, \ W_2=0
%\eeq
\beq 
W_1= W_2=0 \, .
\label{condcomp}
\eeq

In order to write down the conditions for the $3\oplus \bar 3$ 
and $6\oplus \bar 6$ components we need to consider the decomposition 
of the ten-dimensional supersymmetry parameter in terms of the 
four-dimensional one and the SU(3)-invariant spinor.
% and and that these are important ingredients in the 
%remaining equations. 
The two ten-dimensional 
spinors have the same chirality and decompose as 
$\epsilon_i = \zeta_{+} \otimes \eta^i_{+} + \zeta_{-} \otimes 
\eta^i_{-},$ where $i=1,2$ and $\zeta_{+}^{*} =  \zeta_{-}$, 
$\eta_{+}^{i*} = \eta_{-}^i$. 
The six-dimensional spinors $\eta_{\pm}^{i}$ are related to the SU(3) 
invariant spinor $\eta_{+}$ by the functions $\alpha$ and 
$\beta$: $\eta_{+}^1= 
\frac{1}{2} (\alpha +  \beta) \eta_{+}$ and $\eta_{+}^2=
\frac{1}{2i} (\alpha -  \beta) \eta_{+}$, where $\eta_{+}$ is normalized as $\eta_{+}^{\dagger}\eta_{+}=\frac{1}{2}$.
% these define $\alpha$ and  $\beta$. 

For $6 \oplus {\bar 6}$ one gets the following
equations for $W_3$, $F_{3}^{(6)}$ and  $H^{(6)}$
\begin{eqnarray}
&&
(\alpha^2 - \beta^2)\, W_3^{(6)} \,= e^{\phi} \,2 \alpha \beta \,F^{(6)}_{3} \, ,
\nn \\
&& (\alpha^2 + \beta^2)\, W_3^{(6)} \,= - 2 \alpha \beta \,*_6 H^{(6)}
\label{eq:6int} \, , \\
&& (\alpha^2 - \beta^2)\, H^{(6)}=  e^{\phi} \,(\alpha^2 + \beta^2)
*_6F^{(6)}_{3}\nn \, .
\end{eqnarray}
The last equation is related to the self-duality of the complex 3-form flux
$G_3 = F_3 - i e^{-\phi} H$. In our conventions a primitive (1,2) form 
- which transforms in the $6$ representation - is imaginary anti self-dual (AISD), 
while a primitive (2,1) form - transforming in the $\bar 6$ - is ISD.

The analysis for the components 
$3 \oplus {\bar 3}$ is more involved and
depends on the choice of the functions  $\alpha$ and  $\beta$. 
The values $\alpha=0$ or $\beta=0$; $\alpha=\pm \beta$ and $\alpha = \pm i 
\beta$ are special,  since there $W_3$ or $F_{3}^{(6)}$ or $H^{(6)}$ 
vanish. These correspond to well-known cases, which have been labeled
respectively type B, A and C solutions. 
The full set of conditions for the A \cite{Strominger}, B  \cite{GP,Gubser} and C \cite{FG} solutions in a 
compact form are summarized in Table 2 (quantities not mentioned in the table in a given representation are vanishing).
\begin{table}
\begin{center} 
%\hspace{-1.2cm}
\renewcommand{\arraystretch}{1.5}
\begin{tabular}{|c||c|c|c|}\hline 
& $\alpha=\pm\beta$ (A) & $\alpha=0$ or $\beta=0$ (B)& 
$\alpha=\pm i\beta$ (C)  \\ \hline \hline
1&\multicolumn{3}{|c|}{$W_1=F^{(1)}_3=H^{(1)}=0$}\\ \hline
8&\multicolumn{3}{|c|}{$W_2=0$}\\\hline
6&\minicent{2.5}{\vspace{.2cm}$F_3^{(6)}=0$ \\ 
$W_3=\mp * H^{(6)}$\vspace{.2cm}} &
\minicent{3}{$W_3=0$ \\ $\mp e^{\phi} *F_3^{(6)}= H^{(6)}$} &
\minicent{3}{$H^{(6)}=0$ \\ $W_3= \pm e^{\phi} * F_3^{(6)}$} \\ \hline
\multirow{2}{*}{3}& \multirow{2}{*}{
\minicent{2.8}{$ W_5^{(\bar 3)}=2W_4^{(\bar{3})}=\pm 2iH^{(\bar 3)}= 2 \bar\del\phi$\\
$\bar\del A=\bar\del \alpha=0$}
}&
\minicent{4.5}{\vspace{.2cm}$ e^\phi F_5^{(\bar 3)}= 
\frac23 i  W_5^{(\bar 3)} = i W_4^{(\bar{3})}=-2i\bar\partial A =-4i\bar\partial \log \alpha$\\
$\bar \partial \phi =0$ \vspace{.2cm}} &
\multirow{2}{*}{\minicent{4.2}{$\pm e^\phi F_3^{(\bar 3)}=2i  W_5^{(\bar 3)}=$\\
$- 2i\bar\partial A =- 4i\bar\partial \log \alpha=$\\
$-i \bar \partial \phi$}}\\\cline{3-3}
& &
\minicent{4}{\vspace{.2cm}$e^\phi F_1^{(\bar 3)}=2 e^\phi F_5^{(\bar 3)}=$\\ 
$i  W_5^{(\bar 3)} = i W_4^{(\bar{3})}=i\bar\partial  \phi$\vspace{.2cm}} 
& \\
\hline 
\end{tabular}
\end{center}
\begin{center}
Table 2: special IIB solutions with SU(3) structure
\end{center}
\end{table}

\noindent
Type B and C 
include as examples the two known supergravity solutions dual to ${\cal N}=1$
Super Yang-Mills, the KS and MN solutions. 
%KS fits in the type B class of 
%solutions,
%it has a constant dilaton, a conformally Calabi-Yau metric ($W_1=W_2=W_3=0$
%and $3 W_4=2 W_5$) and ISD fluxes ($e^{\phi} F_3=* H$). MN fits instead 
%in the type C class of solutions, it has a complex but not conformally 
%Calabi-Yau metric and non-vanishing $W_5,W_3$ that compensate the RR flux.

%\item{Type B: the Klebanov-Strassler solution\\
The KS solution describes the near horizon geometry of $N$ regular 
and $M$ fractional D3 branes at the tip of the conifold, these 
last being D5 branes wrapping a collapsed 2-cycle of the conifold.
The geometry is the deformed conifold with a 
conformal factor (a ``throat'') due to the presence of fluxes. The topology is 
${\mathbf R} \times S^2 \times S^3$, where at the apex 
of the cone the
$S^2$ shrinks to zero size while the $S^3$ remains finite.\\
The solution falls into the B class of Table 2 for $\beta =0$. 
The dilaton is constant. The metric is conformally Ricci-flat, satisfying  
$W_1=W_2=W_3=0$ and $3W_4 = 2 W_5$.  
The fluxes $H$ and $F_3$
have only non-vanishing components in the $6+\bar 6$, and  
satisfy $e^{\phi} *F_3= H$. Thus the standard NS-RR three-form combination 
$G_3= F_3 - i e^{-\phi} H$ is imaginary-self-dual (ISD).

%\item{ Type C: the Maldacena-Nu\~nez  solution\\
The MN solution describes $N$ D5 branes wrapping the $S^2$ in   
a six-dimensional manifold with topology ${\mathbf R} \times S^2 \times S^3$. 
It is a type C solution with $\beta=  i \alpha$.  
The only non trivial flux is the RR three form which has 
components in the $6+\bar 6$ and
in the $3+\bar 3$. 
The first two components of the intrinsic torsion are zero, $W_1=W_2=0$. The manifold is 
then complex but not conformal Calabi-Yau since $W_4$ is also zero while $W_5$ is not.
$W_5$ is related to the dilaton, the vector component of the RR flux and the warp factor
as in Table 2. $W_3$  matches  the RR flux in the $6$, 
namely $W_3^{(6)}=e^{\phi} * F_3^{(6)}$.

Away from the three special cases, a general set of equations 
for $3 \oplus {\bar 3}$ can be written, but these depend on the phases of 
$\alpha$ and $\beta$. It can be shown that for all Type IIB solutions 
$|\a|^2+|\b|^2=e^A$. Furthermore, 
we can change $\a$ and $\b$ by the same phase with a rotation of the 
spinor $\eta_+$.
A possible choice to describe type B, C and interpolating BC solutions is 
to take $\alpha$ real and $\beta$ imaginary.  
Using this gauge fixing, the resulting set of equations is 
\bea
&& e^{\phi} F^{(\bar 3)}_3 =
\frac{8 \a \beta (\alpha^2 - \beta^2) }{\alpha^4 - 6 \alpha^2 \beta^2 -3  \beta^4}  
\bar \del
\log\alpha \, ,  \nonumber  \\
&& e^{\phi} F^{(\bar 3)}_5 =
 \frac{ -4 i (\alpha^2 +\beta^2)(\alpha^2 - \beta^2) }{\alpha^4 
- 6 \alpha^2 \beta^2 -3  \beta^4}
   \bar \del \log\alpha \, , \label{flux-vector}  \\
&& H^{(\bar 3)} =
 \frac{8i \a\b (\alpha^2 + \beta^2)}{\alpha^4 - 6 \alpha^2 \beta^2 -3  \beta^4}  
\bar \del
\log\alpha \, , \nonumber
\eea
for the fluxes, and 

\begin{tabular}{ll}
\begin{minipage}[t]{6cm}
\bea
&&
W_4^{(\bar{3})} =  \frac{ - 4 (\alpha^2 + \beta^2)^2}{\alpha^4 - 6 \alpha^2 \beta^2 -3  \beta^4}  
\bar \del \log\alpha \, ,  \nonumber \\
&&  W_5^{(\bar 3)} =  \frac{-2( 3\alpha^4 +2\a^2\b^2 + 3\beta^4)}{\alpha^4 - 6 \alpha^2 \beta^2 -3  \beta^4}  \bar \del \log\alpha \, , \nonumber \\
&& \bar \partial  A =  \frac{2 (\alpha^2 - \beta^2)^2}{\alpha^4 - 6 \alpha^2 \beta^2 -3  \beta^4}  \bar \del \log\alpha \, , \quad  \nonumber 
\eea
\end{minipage} \hfill \hfill 
&
\begin{minipage}[t][3.5cm][c]{8cm}
\bea
&&\qquad  \bar \del \phi =   \frac{-16 \alpha^2 \beta^2 }{\alpha^4 - 6 \alpha^2 \beta^2 -3  \beta^4}  \bar \del \log\alpha \,  , \nonumber \\
&& \qquad \bar \del \b =   \frac{-  \beta ( 3\alpha^4 +6\a^2\b^2  -\beta^4) }{\alpha^4 - 6 \alpha^2 \beta^2 -3  \beta^4}  \bar \del \log\alpha \nonumber  , \\
\label{eq:intvec}
\eea
\end{minipage}
\end{tabular}

\vspace{1em}

\noindent for the geometry, 
where, as one can see, all non-vanishing vector components of the 
torsion and the fluxes are proportional to $\bar \del \log\alpha$. 
Notice that only for the three special cases the ratio $\bar \del A / \bar \del \log \alpha$ is constant.

This set of equations seems to suggest the existence of a 
family of supersymmetric type IIB 
solutions with SU(3) structure interpolating between KS and MN.
Finding such a family of backgrounds is the purpose of this  paper. 
At the level of geometry, this involves a family of 
complex metrics that includes the deformed conifold but is not Ricci-flat 
(is not K\"ahler) in general.\footnote{Using the fact that by scaling the fermion one 
can shift away $W_5$ one could relate this to the Hitchin's variational 
problem \cite{Hitchin}- indeed what we have here is an example of a family of closed 
3-forms and the CY metric should come out of the minimization of the 
volume functional. In a way one could speculate that including the 
three-forms $H$ and $F_3$ one could write a ``generalized" Hitchin 
functional 
whose variation should be equivalent to IIB equations of 
motion, and thus our family of backgrounds corresponds to a minimum of 
a volume functional with fluxes.}

\section{Supersymmetry conditions for the interpolating solution}
\label{section3}

We are looking for a family of solutions to the supersymmetry constraints 
(\ref{cond1})-(\ref{eq:intvec}) that interpolates between the KS and 
the MN solutions. 
In \cite{PT} Papadopoulos and Tseytlin  (PT) proposed an ansatz for IIB 
solutions with fluxes involving a generalization of the conifold metric \cite{MT}.
It contains KS and MN  as special cases and it also
describes the singular and resolved conifold metrics.  
This ansatz covers hence solutions in the whole of A-B-C triangle. 
Indeed, as we have already discussed,  KS and MN are
special cases of type B and C respectively, and
by relabeling the fluxes and 
scaling the metric (S-duality), one can 
get the A type from C.   

The idea is then to use PT ansatz to solve the SU(3) structure equations 
given in the previous section. 
The topology of the space described by the ansatz is 
${\mathbf R}^{1,3} \times {\mathbf R} \times S^2 \times S^3$,
where the 6-dimensional metric has $SU(2) \times SU(2)$ symmetry. 
The metric in string frame is given by \footnote{PT use Einstein frame metric in
\cite{PT}, but we found it more convenient to use string frame metric}  
\bea
ds^2 & = &   e^{2 A} dx_{\mu} dx^{\mu} +  e^{-6 p-x}  dt^2  + ds_5^2 =
\nonumber \\
& = &   e^{2 A } dx_{\mu} dx^{\mu} +  \sum_i^6 E_i^2  \, ,
\label{PTmetric}
\eea
where $p$, $x$ and $A$ are functions of the radial coordinate $t$ only.  
We found it more convenient to use slightly different conventions than PT: our radial variable $t$ is
related to the PT one by $du=e^{-4 p}dt$ and similarly the function $2A$ corresponds in PT to $2 p - x +2 A$. 
We also define a new set of vielbeins $E_i$ which are more suitable 
for the SU(3) structure  we will introduce later. 
They are related to the more conventional vielbeins in PT by
\bea \label{vielbeins}
E_1 &=& e^{\frac{x+g}{2}}  e_1 =  e^{\frac{x+g}{2}} d\theta_1 \, ,\nn \\
E_2 & =& e^{\frac{x+g}{2}}  e_2 = -  e^{\frac{x+g}{2}} 
\sin\theta_1 d\phi_1  \, ,\nn \\
E_3 &=& e^{\frac{x-g}{2}}  \tilde{\ee}_1 =  e^{\frac{x-g}{2}}  
(\ee_1 - a(t)  e_1 ) \, , \\
E_4 &=& e^{\frac{x-g}{2}}  \tilde{\ee}_2 =  e^{\frac{x-g}{2}}  
(\ee_2 - a(t)  e_2) \, , \nn  \\
E_5 &=& e^{-3 p - \frac{x}{2}}  dt = e^{p - \frac{x}{2}}  du \, , \nn \\ 
E_6 &=& e^{-3 p - \frac{x}{2}}  \tilde{\ee}_3 = 
e^{-3 p - \frac{x}{2}}  (\ee_3 + \cos\theta_1 d\phi_1) \, , \nn
\eea
where $g$ and $a$ are also functions of the radial coordinate only.
The $S^2$ in the metric is parameterized by the coordinates 
$\theta_1,\phi_1$ and corresponds to the vielbeins 
$e_1$, $e_2$  (their expression can be read off the definition 
of $E_1$ and  $E_2$).
Similarly $\{\ee_1,\ee_2,\ee_3\}$ are the left-invariant forms on  $S^3$
with Euler angle coordinates $\psi,\theta_2,\phi_2$
\bea \label{vielbeins2}
\ee_1 & \equiv & \sin\psi\sin\theta_2 d\phi_2+\cos\psi d\theta_2
\ , \nonumber  \\
\ee_2   &\equiv &  \cos\psi\sin\theta_2 d\phi_2 - \sin\psi d\theta_2
\ ,  \nonumber \\
\ee_3  &\equiv & d\psi + \cos\theta_2 d\phi_2 
\ , \nonumber \\
d \ee_i &=& - \frac{1}{2}\ee_{ijk}  \ee_j \we \ee_k   \ . 
\eea

The fluxes of the PT ansatz respect the $SU(2) \times SU(2)$ symmetry. They
are more readable in the original vielbeins
$e_i$ and $\ee_l$  
\bea
H &=&  h_2  \, \te_3 \we  (\ee_1 \we e_1 + \ee_2 \we e_2 )
+  dt \we \big[ h'_1  (\ee_1 \we \ee_2 + e_1 \we e_2) 
\nonumber \\ 
& & +  \chi' \,  ( e_1 \we e_2 -\ee_1 \we \ee_2)
  + h'_2 \,  (\ee_1 \we e_2 -  \ee_2 \we e_1 )\big]  \, ,\label{hhh}  \\
F_3 & = &  P \, \big[  \te_3\we \big(  \ee_1 \we \ee_2 +  e_1 \we e_2
-  b \,   (\ee_1 \we e_2 - \ee_2 \we e_1) \big) \nonumber \\
& & +  b' \, dt \we (\ee_1 \we e_1 + \ee_2 \we e_2) \big] \, ,   \\
F_5 & =&  {\cal F}_5  +  *{\cal F}_5\ , \\\
{\cal F}_5 & = &  K \,  
e_1 \we e_2 \we \ee_1 \we \ee_2 \we \ee_3 \, . 
\label{fluxes}
\eea
where $h_1$, $h_2$, $b$, $\chi$ and $K$ are function of the coordinate $t$,
and primes  always denote derivatives with respect to $t$. 
The function $K$ is related to $h_1$, $h_2$ and $b$ by
$K(t) = Q + 2 P [h_1(t)+b(t)h_2(t)]$, where the constants 
$Q$ and $P$ are proportional to the number of regular and fractional D3 branes
respectively. In particular, $P=-M\alpha'/4$.

The fluxes and the expression for $K$ are 
chosen in such a way that they automatically satisfy the Bianchi 
identities. This implies that the PT ansatz does not allow 
for the introduction of brane sources and thus always 
describes the ``flux side'' of the 
large N transition \cite{CIV}.

In order to apply the analysis of the previous section to the PT ansatz we 
have to choose an SU(3) structure, which amounts to choosing a fundamental 
form  and a holomorphic 3-form.
\bea \label{cs}
J &=& E_1 \we ( \A  E_2 + \B E_4) +   E_3 \we ( \B  E_2 - \A E_4) 
+ E_5 \we E_6 ,\\
\Omega &= & [E_1 + i ( \A  E_2 + \B E_4)]  \we [ E_3 + i ( \B  E_2 - \A E_4)]
\we [ E_5 + i E_6] \, .
\eea
$\A$ and $\B$ are functions of the radial direction  and correspond to 
a rotation in the $E_2$ - $E_4$ plane (therefore they obey 
$\A^2+\B^2=1$). The introduction of $\A$ and $\B$ is motivated by the fact 
that we need a  structure interpolating between MN and KS.

 This is the structure we will use to decompose the fluxes into  SU(3) 
representations. For the components of the torsions and the 
fluxes in this complex structure, see the Appendix A.

The PT ansatz and the choice of structure  involve the following set 
of unknown
functions: $A$, $a$ , $p$, $x$, $g$ (from  the metric), 
$h_1$, $h_2$, $\chi$, $b$ (from the fluxes), $\A$, 
$\B$, $\alpha$ and $\beta$  (from the SU(3) structure).
The strategy is now to plug the ansatz in equations 
(\ref{cond1})-(\ref{eq:intvec}) and solve the system of differential equations
for our functions.

%All that needs to be done is to take the expressions from the Appendix, 
%substitute into
%equations (\ref{cond1})-(\ref{eq:intvec}) and solve the system.
%First we briefly go over the known solutions.

\subsection{The supersymmetry conditions}

The system of equations for the unknown functions derived 
from (\ref{cond1})-(\ref{eq:intvec}) seems overdetermined and
discouraging. However a lot of patience and a more careful analysis reveals 
that the number of independent equations matches the number of unknowns.
We give here the solution referring to Appendix B for a detailed derivation.
For simplicity, we remove all integration constants 
that can be eliminated with a change of coordinates and those
that correspond to singular backgrounds. The general solution is given in 
Appendix B.

In what follows we also rescale $P=-1/2$.
Inspired by the discussion in Section 2, we  
choose $\alpha$ real and $\beta$
purely imaginary. We can parameterize them as follows
\begin{equation}
\beta= i \tan (w/2) \alpha \, .
\label{par}
\end{equation} 

From the conditions that the complex structure is integrable (\ref{condcomp})
we determine
a functional relation between $a$ and $g$ and the expressions for $\A$ and $\B$
\bea
& &\frac{e^{2 g}+1+a^2}{2 a} = -  \cosh t \, ,\nonumber\\
& & \A=\frac{\cosh t +a}{\sinh t} \, ,\nonumber\\
& & \B=\frac{e^g}{\sinh t} \, .
\label{complex}
\eea

From the scalar, vector and tensor conditions (\ref{cond1})-(\ref{eq:intvec})
we obtain a pair of coupled first-order differential equations
for the quantities $a$ and $v=e^{6p+2x}$ 
\bea
a'=- \frac{{\sqrt{-1 - a^2 - 2\,a\,\cosh t}}\,
\left( 1 + a\,\cosh t \right) }{v \, \sinh t}   - 
  \frac{a\,\sinh t\,\left( t + a\,\sinh t \right) }{t\,\cosh t - \sinh t}, 
\nonumber 
\eea
\beq
\begin{array}{ll}
v' = & \displaystyle \frac{-3\,a\,\sinh t}{{\sqrt{-1 - a^2 - 2\,a\,\cosh t}}} \\[2em]
     & + v \, \left[- a^2 \cosh^3 t + 2\,a\,t \coth t + 
       a \cosh^2 t \left( 2 - 4\,t \coth t \right) + 
       \cosh t \left( 1 + 2\,a^2 \right. \right. \\[1em] 
     & \left. \left. - \left( 2 + a^2 \right) t \coth t \right)
           + t \, \textrm{csch}\,t \right]/ \left[ \left( 1 + a^2 + 2 a \cosh t
       \right) \left( t \cosh t - \sinh t \right) \right] 
\label{coupled}
\end{array}
\eeq
and a set of algebraic and differential equations that allow to
determine all the other unknown functions in terms
of the quantity $a$,
\bea
b &=& -\frac{t}{\sinh t} \, ,\nn\\
h_1&=&h_2 \cosh t \,+Q , \nn\\
h_2' &=& - \frac{\left(t - a^2t + 2at\cosh t + a^2\sinh 2t \right)}
    {\left(1 + a^2 + 2a\cosh t \right) \left(-1 + t\coth t \right)} h_2 \, , \nn\\
%\phi'&=&\frac{-\left({\left(1 + a\cosh t \right)}^2\textrm{csch} t\left(-2t + \sinh 2t \right)\right)}
%  {2\left(1 + a^2 + 2a\cosh t\right)\left( t\cosh t - \sinh t \right)}\nn\\
%x'&=&\frac{2at\cosh t - 2a\sinh t + e^{6p} h_2^2 {\sqrt{-1 - a^2 - 2a\cosh t}}\,
%     \left( -2t + \sinh (2t) \right) }{2e^{2\left(3p + x \right) }{\sqrt{-1 - a^2 - 2a\cosh t}}\left( -1 + t\coth t \right)}\nn\\
%w'&=&-\left(\frac{h_2\left( 1 + a\cosh t \right)\left( \cosh t - t\textrm{csch} t \right) }
%    {e^x{\sqrt{-1 - a^2 - 2a\cosh t}}\left( -1 + t\coth t \right)} \right)\nn\\
\chi'&=&\frac{a h_2\left( 1 + a\cosh t \right) \left( 2t - \sinh 2t \right)}
  {\left( 1 + a^2 + 2a\cosh t \right)\left( -1 + t\coth t \right)} \, , \nn\\
A'&=&-\frac{\left( -1 + t\coth t \right) \left( -\cosh t + t \textrm{csch} t \right)}{8 \, \sinh t}e^{-2x + 2\phi} \, , \nn\\
\sin w &=&- \frac{2e^{x - \phi }\left( 1 + a\cosh t \right) }{{\sqrt{-1 - a^2 - 2a\cosh t}}\,\left( -1 + t\coth t \right) } \, , \nn\\
\cos w &=&\frac{2h_2\sinh t}{e^{\phi }\left( 1 - t\coth t \right) }=\eta e^{\phi}
\label{susyeqs}
\eea
where $\eta$ is an integration constant.

Even if analytically difficult or impossible
to solve, equations (\ref{coupled}) completely determine $a$ and $v$
in terms of two extra integration constants. At this point, equations
(\ref{susyeqs}) allow to determine all the other unknown functions
$x,p,A,w,b,\chi,h_1,h_2$. Finally, the condition $e^A=|\a|^2+|\b|^2$
determines $\a$ and $\b$. The supersymmetric conditions are now completely
satisfied.

\subsection{The MN and KS solutions}

Our solution includes the MN and KS backgrounds as particular cases. 
They correspond to special expressions for the
functions $A$, $a$ , $p$, $x$, $g$, $h_1$, $h_2$, $\chi$, $b$, $\A$, 
$\B$, $\alpha$, $\beta$.

The MN string frame metric  for D5-branes can be recovered for the 
following expressions for the metric functions \footnote{Our solution with these values of parameters coincides with the
MN solution as given in the original paper \cite{MN} when we identify
our radial variable $t$ with $2r$ and change sign to $a$ (which can be done
with a change of coordinates). Recall also that we are writing the metric in the string frame
metric as opposed to Einstein frame used by PT, and thus the functions in the metric
given here look different than those in \cite{PT}.} 
\bea
a=-\frac{t}{\sinh t}\, , \quad \qquad & \qquad  
&e^{2g}=-1+2t\coth t-\frac{t^2}{\sinh^2 t} \, ,\nn\\
e^{2\phi}=e^{2\phi_0}  e^{-g} \, \sinh t \, ,& \qquad & 
e^{2 A}=e^{2 A_0}e^{-g/2} \, \sqrt{\sinh t} \, ,\nn\\
e^x=e^{\phi_0}e^{g/2}\frac{\sqrt{\sinh t}}{2} \, ,
&\qquad &e^{6p}= \frac{4\, e^{-2\phi_0}}{\sinh t} \, ,
\label{MN1}
\eea
and of the fluxes
\beq
h_1 = h_2 = \chi= K=0  \qquad \mbox{and} \qquad a=b \, . 
\label{MN2}
\eeq

The SU(3) structure that gives the results of Table 2 is 
given by (\ref{cs}) for 
\beq
\A= \coth t-t \, \textrm{csch}^2 t, 
\qquad \B=\textrm{csch} t \sqrt{-1+2t\coth t-t^2\textrm{csch}^2 t} \, .
\label{MN3} 
\eeq

It is easy to check that the functions given above solve the 
susy equations ~(\ref{complex})-(\ref{susyeqs}). 
They correspond to a type C solution in Table 2 with
$\beta=i\alpha$, which corresponds to
$w=\pi/2$. 
%In the MN case, the integrated equations ~(\ref{integrated})
%should be used with some care.
%One can check that they also solve the equations \cite{complex}-
%\cite{susyeqs} in the limit where $w, h_2, \eta$ are suitably sent to zero.
%The integration constants should be chosen as $c_1=-1,c_2=0,c_3=0$.

%For completeness, we give the expressions for $W_5$ and $W_3$ 
%\beq
%W_{5 \ ,\bar{i}} = -2 \bar \partial \log \a  =  \frac{-5 (1 - 16 r^2) \cosh 2r - \cosh 6 r + 
%    2 r (5 \sinh 2 r + \sinh 6 r )}{-2^{49/16} sinh^{5/4} 2r  (1 -8 r^2 - \cosh  4 r + 4 r \sinh 4 r )^{15/16}  }
%\eeq
%and
%\bea
%W_3 &=&  e^{\Phi} * F_3^{(6)} \nonumber \\
%&=&   f_1 ( E_1 \wedge E_3 
%-  E_2 \wedge E_4 ) \wedge  E_6  + 
%f_2 ( E_1 \wedge E_2  -  E_3 \wedge E_4 )
%\wedge E_5   \nonumber \\
%& &  + f_3  E_2 \wedge E_3 \wedge E_5     + f_4 E_1 \wedge E_4 \wedge E_5  \, ,
%\eea
%where
%\bea
%f_1&=& \frac{2^{23/16} (- 1+ 2 r \coth 2r )}{ \sinh^{1/4} 2r (1 -8 r^2 - \cosh  4 r + 4 r \sinh 4 r )^{7/16}} \nonumber\\
%f_2 & = &  \frac{2^{15/16} r  (- 4 r + \sinh4r )^2}{ \sinh^6 2r (1 -8 r^2 - \cosh  4 r + 4 r \sinh 4 r )^{15/16}} \nonumber\\\
%f_3&=& \frac{2^{7/16} (- 1+ 8 r^2 + \cosh 4r )}{ \sinh^{9/4} 2r (1 -8 r^2 - \cosh  4 r + 4 r \sinh 4 r )^{7/16}} \nonumber\\\
%f_4 &=& 2^{7/16} \frac{(1 -8 r^2 - \cosh  4 r + 4 r \sinh 4 r )^{9/16}}{ \sinh^{9/4} 2r} \nonumber \, .
%\eea

Similarly the KS solution is obtained by setting in the metric 
\bea
A =-\frac{1}{4} \ln h \, , \quad \qquad \qquad \qquad \, 
& \qquad & a= -\frac{1}{\cosh t} \, , \nn \\
e^{6p+2x}= \frac{3}{2}\, (\coth t-t \, \textrm{csch}^2 t) \, ,  & \qquad &  
e^{g} = \tanh t \, ,  \nn \\
e^{2x}=e^{2\phi_0}\frac{(\sinh t \, \cosh t - t)^{2/3}}{16} h  & \qquad & 
e^{\phi}=e^{\phi_0} \, ,
\label{KS1}
\eea
where
\beq h'= -8 \frac{(t \coth t - 1)(\frac{1}{2}\sinh(2t) - t)^{1/3}}{(\sinh t)^2} \, , 
\eeq
and in the fluxes,
\bea
h_1=\frac{1}{2}(\coth t-t\coth^2 t) e^{\phi_0}
& \qquad &   b= -\frac{t}{\sinh t}  \, , \nn \\
h_2 = -\frac{(-1+t \coth t)}{2\sinh t}  e^{\phi_0}\quad &  \qquad &
\chi=0 \, .
\label{KS2}
\eea
Finally, the SU(3)  complex structure is given by (\ref{cs}) for 
\footnote{This choice of complex structure is related 
to the one used in \cite{PT} by an $SU(3)$ rotations of the vielbeins.}
\beq
\A=e^g ,\quad \B=-a \, .
\eeq

In this case the susy conditions
(\ref{complex})-(\ref{susyeqs}) are identically satisfied with $w=0$ and
$\eta=e^{-\phi_0}$.

Metrically, the deformed conifold is an $S^1$ bundle labeled by the 
coordinate 
$\psi$ in (\ref{vielbeins2}) over $S^2 \times S^2$ parameterized 
by $(\theta_1,\phi_1)$ and $(\theta_2,\phi_2)$. 
%Correspondingly it has an enhanced SU(2) 
%$\times$ SU(2) symmetry, instead of the smaller SU(2) $\times$ U(1) 
%of the interpolating PT ansatz. 
The metric has a $Z_2$ symmetry corresponding to the exchange
$(\theta_1,\phi_1) \rightarrow  (\theta_2,\phi_2)$. This can be seen from the vielbeins 
(\ref{vielbeins})-(\ref{vielbeins2}) and
the fact that $a$ and $g$ in KS are related by $e^{2g}=1-a^2$. Actually, the full solution including
fluxes has an interchange symmetry ${\cal I}$ which is a combination of the  $Z_2$ symmetry 
plus a reversal of the signs of $B_2$ and $C_2$ (the
$-\mathbb I$ of $SL(2,Z)$).  
The MN solution does not have such a symmetry, which is also broken in the  
perturbative solution which we will discuss next.

\subsection{The GHK solution}

A first attempt to go beyond the KS solution was done in \cite{GHK}, where a first order 
deformation of KS was constructed by solving the supergravity
equations of motion.
This deformation  breaks the $Z_2$ symmetry of KS. In terms of the 
PT ansatz (\ref{PTmetric})-(\ref{vielbeins}) it consists
of turning on an additional component in the  NS three-form, $\chi'$,
and of modifying the metric components $a$ and $g$ \footnote{Our $Z$ is 
related to GHK notations by $z_{GHK}=e^{-g}Z$.} 
\beq
a \rightarrow a \left(1 +  Z(t) q \right) \, ,  \qquad
e^g \rightarrow e^g \left(1 + Z(t) q \right) \, ,
\label{GHKd}
\eeq   
where $q$ is the expansion parameter. The $Z_2$ breaking is reflected by the fact that 
the deformation (\ref{GHKd}) does not respect the relation $e^{2g}=1-a^2$ between $a$ and $g$ 
in KS. \\
The other functions are not modified at first order.
The equations of motion fix the function $Z$, and also
relate it to the deformation $\chi'$ in the NS flux:
\bea
Z(t)&=& \frac{(-t + \tanh t )}{(-t + \cosh t \sinh t)^{1/3}} \, ,\nn\\
\chi'(t) &=&-\frac{1}{2}  \coth t \frac{\sinh 2t - 2 t}{\sinh t^2} Z(t) \, . \nn 
\eea
%where $q$ is a (infinitesimal) constant.

The supersymmetry of this solution was not checked in \cite{GHK}. 
With our choice of complex structure it is possible to show that the deformation 
satisfies the susy  conditions (\ref{complex})-(\ref{susyeqs})
with $\a=e^{A/2},\b= -i (q/4) e^{-\phi_0} e^{-3A/2}$. 

It is interesting to examine 
how supersymmetry works in terms of torsions. The flux $G$ is still ISD at first order  
and the metric still conformally Ricci-flat. However, the 
GHK solution has a non-zero $W_3$. This is possible
because $\b$ is deformed at first order. 
The examination of the equations (\ref{cond1})-(\ref{eq:intvec}) shows that 
$W_4,W_5,A$ and $\phi$ do not acquire  a first order
correction. The conformal relation $2W_5=3W_4$ is still satisfied 
but the metric is no longer conformally Calabi-Yau because $W_3\ne 0$.
The reason for Ricci-flatness is that the terms in the Ricci tensor which depend 
on  $W_3$  alone are quadratic, so just like the
self-duality of $G$, the Ricci-flatness of the metric will be violated at 
second order.

\section{A family of IIB backgrounds}
\label{section4}

We wish to go beyond the linear order perturbation of GHK, and find a 
one parameter family of exact IIB backgrounds that
starts from KS and goes up to MN solution, passing through GHK in the 
vicinity of KS. 

The susy  conditions (\ref{complex})-(\ref{susyeqs}) are 
suitable for a perturbative expansion in the deformation parameter $q$. 
For example, we can explicitly solve them at second order in $q$
finding a regular solution.
At this order a perturbation for the functions $A$, $p$, $x$, 
$h_1$, $h_2$, $b$ is turned on. The dilaton also starts to run. 
The expression for $a$ is particularly simple
\beq
a\rightarrow a(1+q Z + q^2 Z^2) \, .
\label{aII}
\eeq
It is unlikely that this simplicity persists at third order.

The perturbative expansion in $q$ is the best we can do analytically.
%{\bf The solution of the coupled equations for $a$ and $v=e^{6p+2x}$ have proved 
%difficult. no clear }  \\
However, the existence of a regular second order solution suggests
that there is indeed a one parameter family of KS deformations. 
This expectation can be confirmed by a  numerical analysis. In the following,
we study the IR and UV asymptotics for the solution and provide
numerical interpolations. Here we anticipate the main results.\\
There is a family of regular solutions that can be parameterized by
the constant appearing in the IR expansion for $a$
\beq 
a=-1+\xi t^2+ O(t^4) \, .
\eeq
$\xi$ ranges in the interval $[1/6,5/6]$, with $\xi=1/2$ corresponding to KS.
The range $[1/2,5/6]$ is related to $[1/6,1/2]$ by the $Z_2$ symmetry.
 All the arbitrary constants in
the supersymmetry equations (except for one arbitrary additive 
constant in the dilaton) are fixed in terms 
of $\xi$ by requiring IR regularity and the absence of an asymptotically
flat region in the UV.
For all values $1/6<\xi\le 1/2$ the solution is asymptotic in the UV to
the KS solution and the dilaton is bounded. 
By fixing the value of the dilaton at $t=0$, we can find a flow between
KS and MN. Indeed, for $\xi\rightarrow 1/6$
the asymptotic suddenly changes, the dilaton blows up in the UV and
the solution smoothly approaches MN. 
%$\xi$ can therefore be interpreted as the parameter of the flow of solutions that interpolate 
%between MN and KS.

%\beq
%\begin{array}{ll}
%v' = & \displaystyle \frac{-3\,a\,\sinh t}{{\sqrt{-1 - a^2 - 2\,a\,\cosh t}}} \\[2em]
%     & + v \, \left[- a^2 \cosh^3 t + 2\,a\,t \coth t + 
%       a \cosh^2 t \left( 2 - 4\,t \coth t \right) + 
%       \cosh t \left( 1 + 2\,a^2 \right. \right. \\[1em] 
%    & \left. \left. - \left( 2 + a^2 \right) t \coth t \right)
%           + t \, \textrm{csch}\,t \right]/ \left[ \left( 1 + a^2 + 2 a \cosh t
%       \right) \left( t \cosh t - \sinh t \right) \right] 
%\label{eqv}
%\end{array}
%\eeq

\subsection{Numerical Analysis of the Family of Solutions}
%We should solve equations ~(\ref{susyeqs}), substituting the values 
%of $b$ and $g$ obtained from  conditions ~(\ref{complex}),(\ref{susyeqs}). 
%We obtain a first order system of two differential equations in the variables 
%$a, v$; we were not able to find an explicit form for a generic interpolating 
%solution between KS and MN, but we showed that such regular solutions exist through an analysis
%based on power series expansions near $t=0$. We studied also the asymptotic 
%behavior of the solutions for $t\rightarrow \infty $ and then performed a 
%numerical analysis to check our results.
%\subsection{Integration of the susy equations by series}
\label{expansions}

As already mentioned, we were not able to find analytical solutions to the system
of equations ~(\ref{coupled}). However it is possible to show that a one 
parameter family of regular solutions exists by performing a 
power series expansion near $t=0$. In this Section we discuss some details of this analysis.

%By solving the system of equations ~(\ref{susyeqs}) by series, we find a one 
%parameter family of regular solutions. 

Let us consider first the IR behavior of the solution. 
We calculated the first fourteen terms of the expansion around $t=0$ 
for $a$ and $v$; the results up to order $t^4$ are 
\bea
a & = & -1 + \xi \,t^2 + \frac{\left( -3 + 29\,\xi  - 114\,{\xi }^2 + 36\,{\xi }^3 \right) \,t^4}{60} + O(t^6) \, ,\\[0.5em]
%  & + & \frac{\left( -555 + 11762\,\xi  - 76800\,{\xi }^2 + 198720\,{\xi }^3 - 118800\,{\xi }^4 + 23328\,{\xi }^5
%  \right) \,t^6}{50400} + O(t^7) \, ,\nn \\[1em]
v & = & t + \frac{\left( 5 - 84\,\xi  + 84\,{\xi }^2 \right) \,t^3}{120} + O(t^5) \, .
% & + & \left( 439 - 4920\,\xi  + 21672\,{\xi }^2 - 33504\,{\xi }^3 + 16752\,{\xi }^4 \right) \,t^5 /(13440) +  
%  O(t^7) \nn
\eea   

We could expect a two-parameter
space of solutions of the system ~(\ref{coupled}). 
However, one can see that for one of the two parameters the solutions 
are not regular in $t=0$ \footnote{See also eq 5.75 in ref.(\cite{GHK}).}.
%, or, starting from an arbitrary $t$, does not reach $t=0$. 
Notice also  that $a(0)=-1$ and $v(0)=0$ for any regular solution.
$\xi$ parameterizes the family of solutions of the system ~(\ref{coupled}), 
with $\xi =1/2$ and  $\xi=1/6$ corresponding to the KS and MN solutions, respectively.
A numerical analysis shows that the solutions exist and are regular for $1/6 \leq \xi \leq 1/2$.
%$\xi =1/2$ corresponds to the KS solution and $\xi=1/6$ corresponds to the MN solution. 
%$\xi$ can therefore be interpreted as the parameter of the flux of solutions that interpolate 
%between MN and KS, and we expect that such solutions exist and are regular, at least for 
%not too large values of $t$. Numerical integration has confirmed the existence of these 
%interpolating solutions for every value of $t>0$ when $\xi$ varies 
%in the range $1/6 \leq \xi \leq 1/2$, that is from MN to KS. 
Moreover for all the solutions in this range $a\rightarrow 0$ for $t\rightarrow \infty$. 
Actually we will see in the next section that, because of a $Z_2$ symmetry around $\xi=1/2$, 
the flow of solutions exists for $1/6 \leq \xi \leq 5/6$. Outside this range of values 
we did not find (with numerical simulations) a regular solution surviving for every $t>0$. 
% in the point $t=0$ there is no uniqueness for the solutions; instead for $t>0$ the 
%vector field we are integrating is regular in an opportune range of values for $a$ and $v$. 
%The integration parameter is $\xi$, proportional to the second derivative of $a$ in $t=0$. 

Knowing the series expansions for $a$ and $v$, we can use the conditions 
~(\ref{susyeqs}) to determine the other unknown functions.
%we have now to perform two other integrations: the equation for the dilaton $\phi$ and 
%the equation for $x$, then we can use the algebraic rules to deduce $h2$ and $w$. 
%Alternatively one could integrate the equations for $\phi$ and $h2$ and deduce from 
%the algebraic rules $x$ and $w$. The calculations of other expressions are straightforward. 
We list for every quantity the first two non-zero terms in the resulting series:
\bea
& & \phi =  \phi_0 + \frac{\left( 1 - 2\,\xi  \right)^2 \,t^2}{4} + 
  \frac{\left( 1 - 2\,\xi  \right)^2 \,\left( 13 - 132\,\xi  
+ 132\,{\xi }^2 \right) \,t^4}{480} + O(t^6) \, , \\
& & e^x  =  \frac{e^{\phi_0}\,\lambda \,t}{2} + e^{\phi_0} \frac{
     \left( -40 + 3\,\left( 35 - 108\,\xi  
+ 108\,{\xi }^2 \right) \,{\lambda }^2 \right) \,t^3}{720\, \lambda } + O(t^5) \, ,\\
& & h_2 =  e^{\phi_0} \sqrt{4 - 9\,{\left( 1 - 2\,\xi  \right) }^2\,{\lambda }^2} 
\left[ \frac{-t}{12} + \frac{\left( -2 + 15\,\xi  - 15\,{\xi }^2 \right) \,t^3}{90} 
+ O(t^5) \right] , \label{serieh2} \\
& & \sin w  =   \frac{\left( 3 - 6\,\xi  \right) \,\lambda }{2} - 
  \frac{\left( -1 + 2\,\xi  \right) \,
\left( -4 + 9\,{\left( 1 - 2\,\xi  \right) }^2\,{\lambda }^2 \right) \,t^2}{24\,\lambda } 
+ O(t^4) \, , \\
& & A  =  A_0 + \frac{t^2}{18\,{\lambda }^2} + 
  \frac{\left( 40 + \left( -51 - 36\,\xi  + 36\,{\xi }^2 \right) \,{\lambda }^2 \right) \,t^4}
   {6480\,{\lambda }^4} + O(t^6) \, .  
\eea
The expressions for $b$, $g$, $h_1$, $\chi'$ can be obtained by the algebraic 
rules (\ref{complex}),(\ref{susyeqs}).

The IR solution depends on three independent integration constants: $\xi$, $\phi_0$ and $\lambda$.
Indeed the constant $A_0$ can be reabsorbed in a 
rescaling of the space time coordinates $x_m$ in $e^{2A} dx_m^2$. 
The constant $\phi_0$ can take any real value and corresponds to the value of the dilaton 
in $t=0$. The third parameter, $\lambda$, 
describes the behavior of $e^x$ near $t=0$ and it determines
the radius of the IR $S^3$. 
Its role in the gauge theory is explained in Section 5. $\lambda\sim 0.93266$ for KS and $\lambda=1$ for MN
\footnote{The range of $\lambda$ should be determined 
by imposing that $|\sin w|$ and $|\cos w|$ are 
less than 1 for every $t$ and that the exponentials $e^x$ or $e^p$ are always positive.}. 
%For instance in $\xi=1/6$ (MN), where we know the exact solution for $a$ and $v$, 
%one can show by a direct integration that $\lambda$ must be equal to 1; for generic 
%values of $\xi$ in $(1/6,5/6)$, $\lambda$ ranges in an interval of values close to 1.
Notice also that the quantities $a$, $v$, $b$, $g$ do not depend on $\phi_0$ and $\lambda$, 
and $h_2$ depends on these parameters only through an overall  multiplicative factor.

We now  study the susy equations near $t \rightarrow \infty$. 
This can be done by expanding all functions 
in our ansatz in powers of $e^{-t/3}$ times polynomial coefficients in $t$ \cite{apreda}. 
The results for the system $(a,v)$ are: 
\bea
a & = & -2 e^{-t} + a_{UV}\,\left( -1 + t \right) e^{-\frac{5\,t}{3}} - 
 \frac{1}{2}\,{a_{UV}}^2\, {\left( -1 + t \right) }^2 \,e^{-\frac{7\,t}{3}} 
+ O(e^{-3t}) \, ,\label{infa}  \\
v & = & \frac{3}{2} + \frac{9}{16}\,{a_{UV}}^2\,e^{-\frac{4\,t}{3}}\,
\left( 6 - 4\,t + t^2 \right) 
    - \frac{3}{32}\,e^{-2\,t}\,\left( 33\,{a_{UV}}^3 + 
       16\,\left( 7 - 3\,c + 4\,t \right)  \right) + O(e^{-\frac{8\,t}{3}}) \, .\nn 
\eea  

The UV behaviors are parametrized by $a_{UV}$. The second integration constant for $a$ and $v$,
called $c$ \footnote{We find it for the first time in $v$ in the coefficient 
of $e^{-2\,t}$ and in $a$ in the coefficient of $e^{-\frac{11\, t}{3}}$}, can be considered as a function of $a_{UV}$: 
$c=c(a_{UV})$.
Indeed we know that there is only a one parameter family of solutions regular in $t=0$.
% the coefficient $c$ must be chosen in function of $a_i$ in order to be able 
%to continue regularly the solution until $t=0$. 
%So we consider $c$ as a function of $a_i$: $c=c(a_i)$. 
The  parameter $a_{UV}$ is the ultra-violet analogue of $\xi$: it drives the solutions from KS 
to MN, corresponding to $a_{UV}=0$ and $a_{UV}=-\infty$, 
respectively\footnote{In fact (\ref{infa}) does not reproduce the ultraviolet behavior 
for the MN solution, which 
for $a$ reads $a=-2te^{-t}+O(e^{-3\,t})$. The system ~(\ref{coupled})
admits another solution in power series for large $t$ which corresponds to MN.
This second solution does not contain any 
arbitrary integration constant.}. 
%In fact the ultraviolet behavior for the MN solution cannot be expressed in the form given 
%in (\ref{infa}) since the expansions of $a$ for MN is $a=-2te^{-t}+O(e^{-3\,t})$: 
%when solving for large $t$ the second order equation for $a$ coming from the 
%system (\ref{coupled}), one finds two kinds of solutions: one is 
%reported in (\ref{infa}), the other is the MN solution (no arbitrary integration 
%constant appears in this latter case). 

The first terms in the large $t$ expansion of the other quantities read 
\bea
\phi & = & \phi_{UV}+\frac{3}{64}\,{a_{UV}}^2\,e^{-\frac{4\,t}{3}}\,
\left( 1 - 4\,t \right) 
 + O(e^{-\frac{8\,t}{3}}) \, ,\\
h_2 & = & d \left[ e^{-t}\,\left( 1 - t \right)  
+ \frac{3}{32}\, {a_{UV}}^2 \,e^{-\frac{7\,t}{3}}\,
\left( -1 + t \right) \,\left( -1 + 4\,t \right) + O(e^{-3\,t}) \right] \, ,\\
e^{2x} & = & \frac{e^{2\, \phi_{UV}}-d^2}{{a_{UV}}^2} \, 
e^{\frac{4\,t}{3}} \label{infe2x} \\
  & &  + \, \frac{1}{32}\left(-2\,d^2\,{\left( 5 - 2\,t \right) }^2 
+ e^{2\,\phi_{UV}}\,
   \left( 47 - 28\,t + 8\,t^2 \right)\right)+O(e^{-\frac{2\,t}{3}}) \, ,\nn \\
\cos w & = & d\,e^{-\phi_{UV}} \left[ 1 - \frac{3}{64}\,{a_{UV}}^2\,e^{-\frac{4\,t}{3}}\,
\left( -1 + 4\,t \right)  \right] + O(e^{-\frac{8\,t}{3}}) \, , \\
e^{2A}& = &e^{2A_1}\left[1-
\frac{3 {a_{UV}}^2 e^{2\,\phi_{UV}}}{64 \left( e^{2\,\phi_{UV}}-d^2 \right)}
(4t-1)e^{-\frac{4\,t}{3}} + O(e^{-\frac{8\,t}{3}}) \right] \, .
\eea

%These asymptotics depend on the constants $\varphi$ and $d$.
The other two UV integration constants are $\phi_{UV}$ and $d$.
The first is the UV value of the dilaton. If we fix $\phi_0$,
$\phi_{UV}(\xi)$ 
is a function of the flow parameter $\xi$, which
 can be found by numerical analysis.
%is the constant approached for large $t$ by the dilaton if we set to zero $\phi(0)=\phi_0$ and
% so it is a function of the flux parameter $a_i$. 
Obviously $\phi_{UV}(1/2)=\phi_0$ in the KS case; numerical estimates of 
$\phi_{UV}(\xi)$ 
indicate that it diverges approaching MN. 
The second constant $d$ appears multiplicatively in the equation for $h_2$: it is 
the ultraviolet analogue of $\lambda$, or better of a combination of $\lambda$ and $\xi$. 

The UV integration constants can be given in terms of the IR ones by 
matching the solution 
from the regions near $t=0$ and near $t \rightarrow \infty$.
One can for instance find the relation 
between $a_{UV}$ and $\xi$. This can be done numerically and it shows that 
$a_{UV} \rightarrow -\infty$ when $\xi \rightarrow 1/6$. 
The interval $1/6 \leq \xi \leq 1/2$ gets mapped into $-\infty < a_{UV} \leq 0$ 
and this also shows that for $\xi<1/6$ one should not expect regular solutions 
approaching zero for $t\rightarrow \infty$ as in (\ref{infa}). 
$a_{UV}$ can also assume positive values, which correspond to $1/2 \leq \xi \leq 5/6$, 
since the $Z_2$ symmetry acts as $a_{UV} \rightarrow -a_{UV}$ (see next section).   
Similarly one can determine the behavior of $d(\xi,\lambda)$. 

In fact 
we can do more and deduce an exact relation between the IR and UV parameters. 
We know that $e^{\phi} /\cos w $ is a constant for every value of $t$ (see (\ref{susyeqs})). Thus 
equating the two constants obtained by expanding the expression $e^{\phi} /\cos w $ 
for small $t$ and for large $t$ we get:
\beq
d\,e^{-2\,\phi_{UV}}= e^{-\phi_0} {\sqrt{1 - \frac{9\,{\left( 1 - 2\,\xi  \right) }^2\,{\lambda }^2}{4}}}
\label{lambda}
\eeq 
In the notations of equation ~(\ref{susyeqs}), $\eta=d \, e^{-2\phi_{UV}}$.

At this level we still have a solution labeled by two independent parameters (plus the parameter for the dilaton). 
However,
for generic values of $d$ and $\phi_{UV}$, $e^{2x}$ diverges exponentially
and, moreover,
$e^{2A}$ approaches a 
constant for $t\rightarrow \infty$.
In order to eliminate the asymptotic minkowskian region we fix 
$d=e^{\phi_{UV}}$.
This is a requirement from AdS/CFT correspondence, since we want our supergravity solution 
to have a boundary for $t \rightarrow \infty$.
The UV expansion is now, up to an arbitrary multiplicative constant:
\beq
e^{2A} = \frac{e^{\frac{2\,t}{3}}}{{\sqrt{-1 + 4\,t}}} + \frac{{a_{UV}}^2\,
     \left( -847 + 1752\,t - 864\,t^2 + 256\,t^3 \right) }{1024\,
     {\left( -1 + 4\,t \right) }^{\frac{3}{2}}} e^{-\frac{2\,t}{3}} + O(e^{-\frac{4\,t}{3}})
\eeq

The expression (\ref{lambda}), for $d=e^{\phi_{UV}}$, 
allows also to determine $\lambda$ in terms of $\xi$.
For $\xi\rightarrow 1/6$ or 
$\xi \rightarrow 5/6$, $e^{\phi_{UV}}$ diverges. 
This implies that $\lambda$ approaches 1, which is the value for MN.

In  Figure 1 we plot the behavior of the functions $a,\phi, h_2$ and $e^x$ 
as a function of $t$ for several values of $\xi$ and in 
Figure 2 we plot the behavior of the UV values $a_{UV}$ and $\phi_{UV}$ as a 
function of $\xi$. In these plots we have fixed $\phi_0$. 
Notice that the dilaton is always bounded except for the 
values $\xi\rightarrow 1/6$ or $\xi \rightarrow 5/6$ where it diverges at
large $t$.

\begin{figure}[t]
\begin{minipage}[t]{\linewidth}
~~~\begin{minipage}[t]{0.5\linewidth}
\vspace{0pt}
\centering
\includegraphics[scale=0.7]{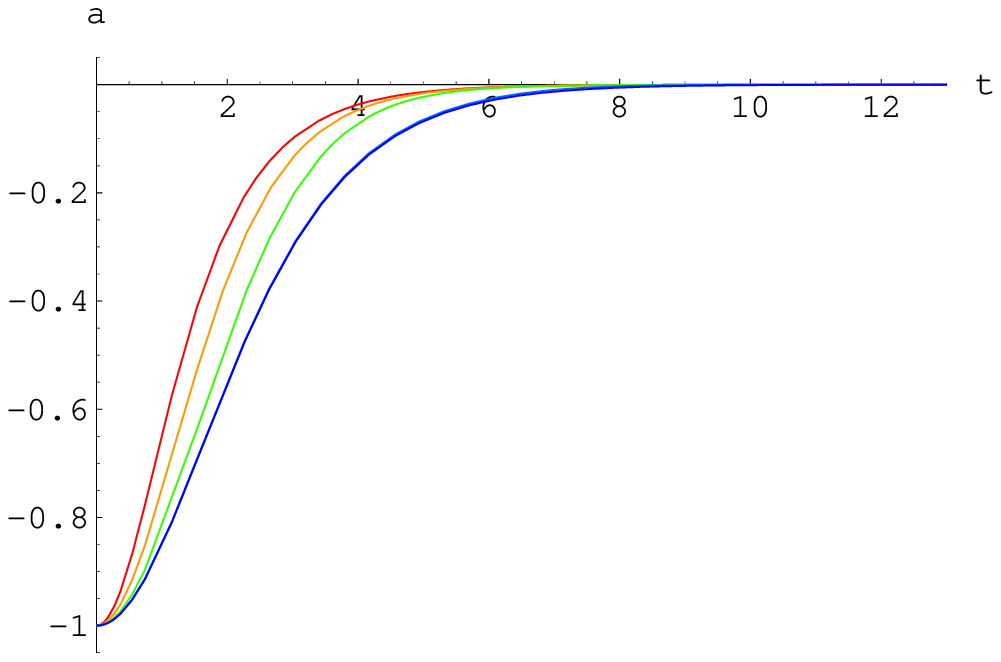}
\end{minipage}%
\begin{minipage}[t]{0.5\linewidth}
\vspace{0pt}
\centering
\includegraphics[scale=0.7]{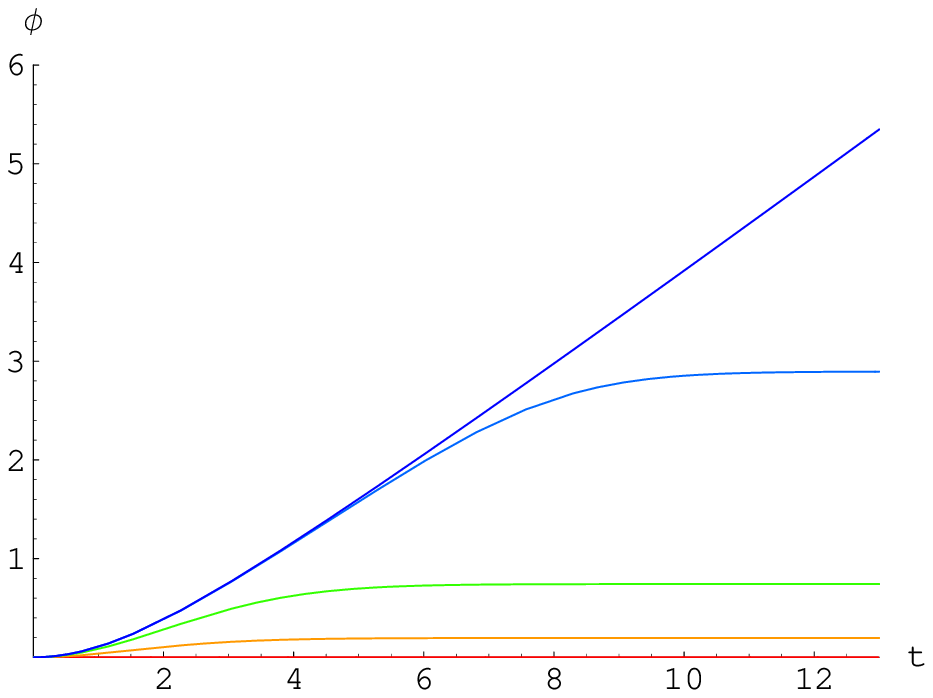}
\end{minipage}\\[1em]
\begin{minipage}[t]{0.5\linewidth}
\vspace{0pt}
\centering
\includegraphics[scale=0.7]{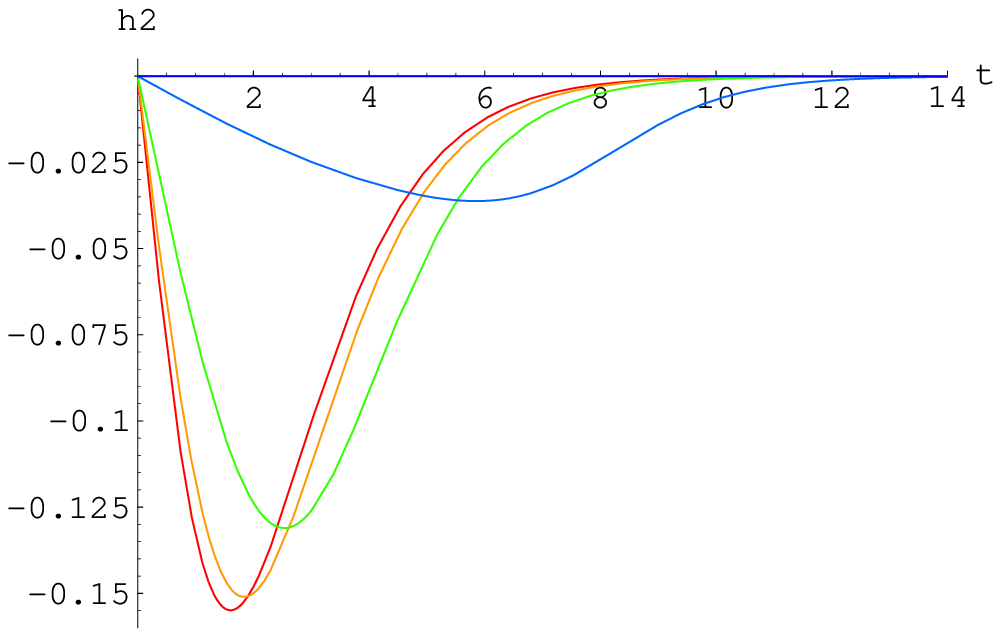}
\end{minipage}%
\begin{minipage}[t]{0.5\linewidth}
\vspace{0pt}
\centering
\includegraphics[scale=0.7]{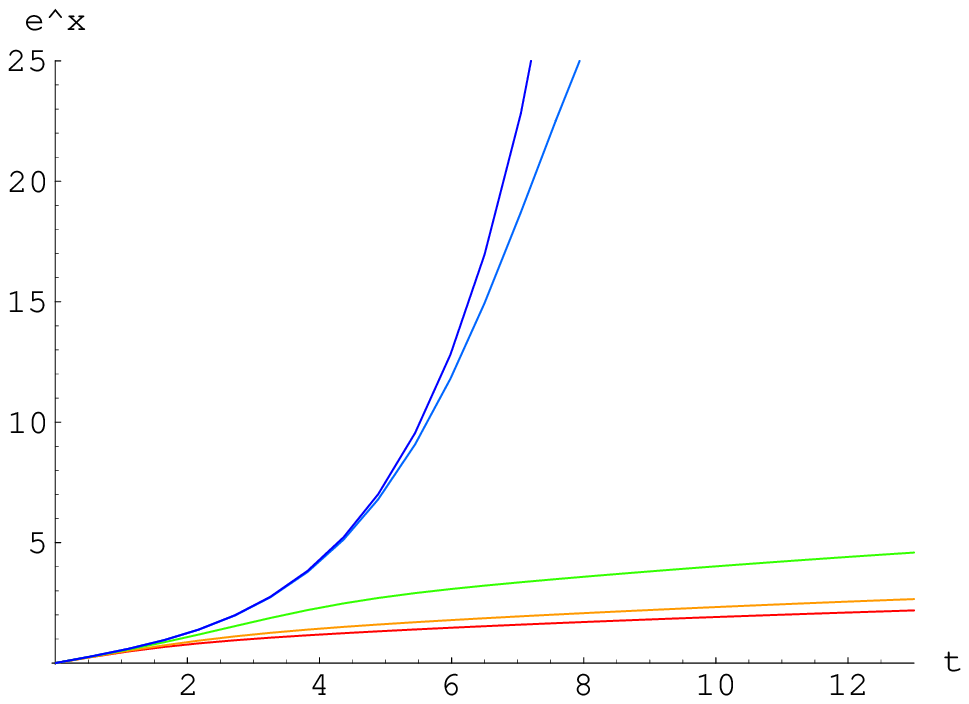}
\end{minipage}%

\vspace{1em}
Figure 1: graphics of the functions $a$, $\phi$, $h_2$ and $e^x$ for different values of the flow parameter $\xi$: the red curves correspond to KS solution ($\xi=1/2$), the blue curves to MN solution ($\xi=1/6$) and the orange, green and light blue curves correspond respectively to the interpolating values $\xi=0.3$, $\xi=0.2$ and $\xi=0.167$. 
\end{minipage}
\end{figure}

\begin{figure}[t]
\begin{minipage}[t]{\linewidth}
~~~\begin{minipage}[t]{0.5\linewidth}
\vspace{0pt}
\centering
\includegraphics[scale=0.7]{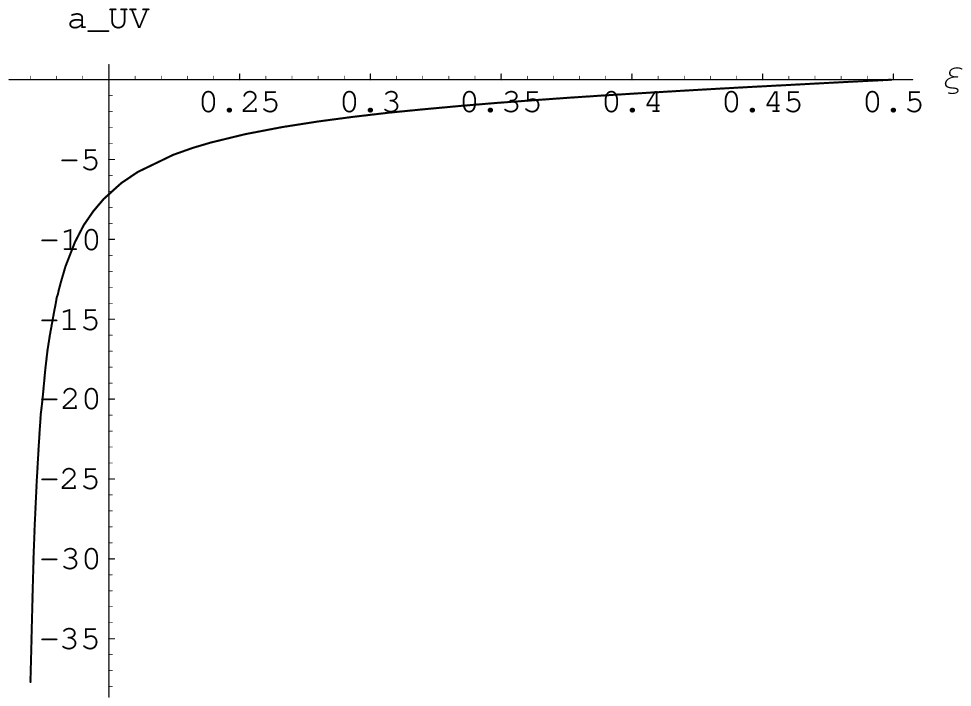}
\end{minipage}%
\begin{minipage}[t]{0.5\linewidth}
\vspace{0pt}
\centering
\includegraphics[scale=0.7]{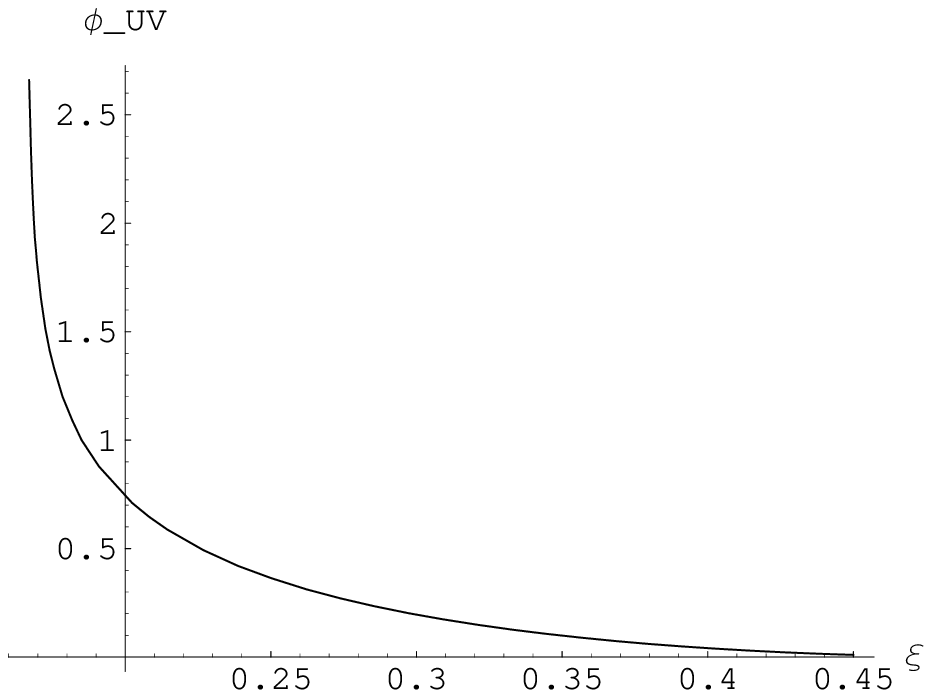}
\end{minipage}\\[1em]
\begin{minipage}[t]{0.5\linewidth}
\vspace{0pt}
\centering
\end{minipage}%

\vspace{1em}
Figure 2: graphics of $a_{UV}$ and $\phi$ as a function of the flow parameter $\xi$, for the fundamental region
$1/6 \le \xi \le 1/2$.
KS solution corresponds $\xi=1/2$ while MN to $\xi=1/6$. 
\end{minipage}
\end{figure}

\subsection{The $Z_2$ symmetry}

As anticipated in the previous section, our susy equations have a $Z_2$ symmetry: 
for any interpolating solution $(a,v)$ of the system (\ref{coupled}), we have 
another solution $(\tilde{a},v)$ with the same function $v\equiv e^{2x+6p}$, but, in general, 
with a different form for $a$. In this Section we will show that $Z_2$ acts on 
the flow parameter $\xi$ as $\xi\rightarrow (1-\xi)$ and on the UV parameter 
$a_{UV}$ as $a_{UV}\rightarrow -a_{UV}$. Note that the KS solution $\xi=1/2$ or $a_{UV}=0$ is invariant 
under $Z_2$, while as soon as we move from KS along the flow of solutions, the $Z_2$ 
symmetry is broken.

To prove such a symmetry one can introduce the following expression
\beq
m\equiv a\,e^{-g} = \frac{a}{\sqrt{-1 - a^2 - 2\,a\,\cosh t}}
\label{relam}
\eeq
where we used the explicit algebraic relation between $g$ and $a$. 
The system (\ref{coupled}) can be unambiguously rewritten in terms of the 
new variables $(m,v)$:
\bea
m' & = & \frac{ -1 - m^2 + m^2\,\cosh^2 t }{v \,\sinh t} + 
  \frac{m\,\sinh t\,\left( t + m^2\,t - m^2\,\cosh t\,\sinh t \right) }
   {- t\,\cosh t  + \sinh t} \, , \label{eqm} \\[0.5em]
v' & = &  \frac{  m^2\, \cosh^3 t - 
       \cosh t\,\left( -1 + m^2 + 
          \left( 2 + m^2 \right) \,t\,\coth t \right)  + 
       \left( 1 + m^2 \right) \,t\,\textrm{csch}\,t }{-\sinh t + t\,\cosh t}\,v \nn\\ & & - 
  3\,m\,\sinh t  \, .
\label{vfm}
\eea

Inverting the relation (\ref{relam}) between $a$ and $m$ we find a second order equation with the following solutions:
\beq
a_{\pm}=\frac{- m^2\,\cosh t \pm  \sqrt{-m^2 - m^4 + m^4\,\cosh^2 t}}{1 + m^2} \, .
\label{apm}
\eeq
Therefore given a solution $(m,v)$ for the system (\ref{eqm}), (\ref{vfm}), we obtain two 
solutions for (\ref{coupled}) with the same function $v$: $(a_{+},v)$ and $(a_{-},v)$. 
These are the two solutions connected by the $Z_2$ symmetry.
A power series solution around $t=0$ of (\ref{eqm}), (\ref{vfm})  shows that
 $a_+\sim-1+\xi t^2$ and $a_-\sim-1+(1-\xi)t^2$. We see therefore that the 
$Z_2$ action is $\xi\rightarrow (1-\xi)$.
%: if one solves (\ref{eqm}), (\ref{vfm}) expanding in series near $t=0$, then can 
%check that $a_{\pm}$ correspond to $a\sim-1+\xi t^2$ and $a\sim-1+(1-\xi)t^2$; 
Similarly one can see that under $Z_2$  $a_{UV}\rightarrow -a_{UV}$.
\footnote{It is also possible to show that the constant $c$ in the expansion for $v$ has the form $c=3+11/16 \, a_{UV}^3+f_p(a_{UV})$, 
where $f_p$ is an even function $f_p(a_{UV})=f_p(-a_{UV})$ and $f_p(0)=0$ 
(the coefficient 3 is determined by a comparison with the first order GHK deformation).} 

So we have shown that the combination $m=a\,e^{-g}$ is invariant under $Z_2$.
In the same way, replacing $a$ with $m$ in the equations for $\phi$, $h_2$, $x$, $p$ and $A$ 
it is easy to prove that also these quantities are  $Z_2$ invariant.
%one can rewrite the equations for $\phi$, $h_2$, $x$ and $A$ replacing $a$ with $m$ 
%(\ref{relam}) showing that also these other quantities are $Z_2$ invariant: 
%$\phi \rightarrow \phi$, $h2\rightarrow h2$, $x\rightarrow x$, $p\rightarrow p$, 
%$A\rightarrow A$, $b\rightarrow b$. 
On the contrary $\chi'\rightarrow -\chi'$. 

Since we know the exact form for $a$ in the MN case $\xi=1/6$,  one can deduce from (\ref{relam}), 
(\ref{apm}) the exact solution when $\xi=5/6$: we find 
$a=t/(-2t\cosh t+\sinh t)$. Note also that while in the MN solution the 
angle $w$ is equal to $\pi/2$, its $Z_2$ symmetric solution has $w=-\pi/2$.

Now we want to relate the $Z_2$ symmetry of the susy equations to the $Z_2$ symmetry of 
metric\cite{GHK} corresponding  to the exchange of  the two spheres $(\theta_1,\phi_1)\leftrightarrow(\theta_2,\phi_2)$. 
The PT metric ansatz can be written as
\beq
\begin{array}{lll}
ds^2 & = & e^{2A}dx_m^2+e^x(e^g+a^2e^{-g})(e_1^2+e_2^2)+e^{x-g}(\epsilon_1^2+\epsilon_2^2)-2ae^{x-g}(e_1\epsilon_1+e_2\epsilon_2) \nn \\[0.5em]
     &   & + e^{-6p-x}(\tilde{\epsilon_3}^2+dt^2)
\end{array}
\eeq
It is easy to check from (\ref{relam}), (\ref{apm}) that under $Z_2$: $e^g+a^2e^{-g}\leftrightarrow e^{-g}$ thus exchanging the coefficients in front of the two $S_2$. 
In the same way one can check that the fields transform under $Z_2$ as: $F_3 \rightarrow -F_3$, $H_3 \rightarrow -H_3$, $F_5 \rightarrow F_5$. This is just the action described in \cite{GHK}.

Since the $Z_2$ symmetry is implemented by a change of coordinates,
two vacua related by a $Z_2$ transformation are equivalent.

\section{Gauge theory}
\label{section5}

The gauge theory dual to the supergravity background corresponding to the KS solution was studied in \cite{KS}, where it was identified as the gauge theory on a stack of $N$ regular and $M$ fractional D3-brane at the apex of a conifold. The resulting theory is a $\mathcal{N}=1$ susy $SU(N+M)\times SU(N)$ gauge theory with chiral fields $A_i$, $B_j$ respectively in the $(N+M,\overline{N})$ and $(\overline{N+M},N)$ of the gauge group, transforming as doublets of the $SU(2)\times SU(2)$ global symmetry group. 
The theory undergoes repeated Seiberg-duality transformations in which $N\rightarrow N-M$, until in the far infrared the gauge group is reduced to $SU(M+p)\times SU(p)$, with $0 \leq p <M$. The supergravity background in \cite{KS}, like  the interpolating backgrounds in this paper, corresponds to $p=0$ since the field $F_5$ approaches zero for $t=0$; in this case the supergravity solution is regular in the IR and reliable for large values of the 't Hooft coupling $g_s M$; the factor $e^{2A}$ is constant for $t=0$ and consequently the gauge theory is confining, as shown in \cite{KS}; moreover it has other interesting features such as chiral symmetry breaking $Z_{2M}\rightarrow Z_2$ via gluino condensate, domain walls, magnetic screening. 

When $p=0$ the gauge group in the far IR is $SU(M)$, and as suggested in \cite{aharony,KS}, the last step in the duality cascade ($SU(2M)\times SU(M)$) is on the baryonic branch, i.e. the $U(1)_B$ global symmetry ($A_i\rightarrow e^{i\alpha}A_i$, $B_j\rightarrow e^{-i\alpha} B_j$) is broken by the expectation values of baryonic operators:
\bea
\mathcal{B} \sim \epsilon_{\alpha_1 \alpha_2 \ldots \alpha_{2M}} (A_1)_1^{\alpha_1} (A_1)_2^{\alpha_2}\ldots (A_1)_M^{\alpha_M} (A_2)_1^{\alpha_{M+1}} (A_2)_2^{\alpha_{M+2}}\ldots (A_2)_M^{\alpha_{2M}} \\
\bar{\mathcal{B}} \sim \epsilon^{\alpha_1 \alpha_2 \ldots \alpha_{2M}} (B_1)^1_{\alpha_1} (B_1)^2_{\alpha_2}\ldots (B_1)^M_{\alpha_M} (B_2)^1_{\alpha_{M+1}} (B_2)^2_{\alpha_{M+2}}\ldots (B_2)^M_{\alpha_{2M}} 
\eea
The baryonic branch has complex dimension 1, and it can be parametrized by $\zeta$:
\beq
\mathcal{B}=i \zeta \Lambda_{2M}^{2M}, \qquad \bar{\mathcal{B}}=\frac{i}{\zeta} \Lambda_{2M}^{2M}
\eeq
where $\Lambda_{2M}$ is the UV scale of the gauge group $SU(2M)$. Note that the $U(1)_B$ corresponds to changing $\zeta$ by a phase.
The Goldstone boson associated to the spontaneous breaking of the $U(1)_B$ symmetry was identified in the supergravity dual as a massless pseudo-scalar bound state (glueball) in \cite{GHK}, where it was also suggested that D1-branes in string theory are dual to axionic strings in gauge theory that create a monodromy for this massless axion field.    

By supersymmetry the Goldstone boson is in a $\mathcal{N}=1$ chiral multiplet; hence there will be a massless scalar mode, the ``saxion'' that must correspond to changing $\zeta$ by a positive real factor. This is a modulus of the theory whose expectation value induces a one parameter family of supersymmetric deformations. In the same paper \cite{GHK}, the supergravity dual of such deformations was suggested to be the GHK solution, constructed to first order in the flow parameter. We have shown that such deformation is supersymmetric (since it satisfies the susy equations we wrote), and moreover that a supersymmetric extension of it exists to all orders.

In \cite{GHK}, the saxion operator in the gauge theory was identified as:
\beq
\textrm{Re} \, \textrm{tr} \left[ a_i^{*} \square a_i- b_i^{*} \square b_i \right] + \textrm{fermion bilinears}.
\label{saxion}
\eeq
Where $a_i$ and $b_i$ are the lowest components of the chiral fields $A_i$ and $B_i$. This operator is odd under the $Z_2$ symmetry discussed in the previous section, which corresponds in field theory to the interchange of $A_1$, $A_2$ with $B_1$, $B_2$ (see \cite{GHK}) accompanied by the charge coniugation. The KS solution, being invariant under $Z_2$, corresponds to a vacuum where $|\mathcal{B}|=|\bar{\mathcal{B}}|=\Lambda_{2M}^{2M}$, and giving a non zero vev to the saxion operator (\ref{saxion}), that is perturbing $|\zeta|$, should be the gauge theory analogue of our $Z_2$ breaking supergravity solution. 

A word of clarification is due about the uses of ``interpolation"  and
``baryonic branch". The solution constructed in this paper depends on
two parameters, $\xi$ and the additive constant in the dilaton.
In the previous Section we constructed an interpolation between KS and MN 
(type B and C). This interpolation involves fixing the IR value of the 
dilaton $\phi_0$. This necessarily implies that the UV value
of the dilaton $\phi_{UV}$ varies along the flow. 
On the other hand, when discussing the dual gauge theory
one has to fix the UV value of the dilaton \footnote {We thank A. Dymarsky and I. Klebanov for pointing this
out.}. What one calls the baryonic branch corresponds to varying
$\xi$ while keeping $\phi_{UV}$ fixed; otherwise in addition to changing
baryonic VEVs we would be changing the parameters of the Lagrangian also.

We therefore suggest that the family of deformations we have found, when
$\phi_{UV}$ is fixed, describes different vacua of the same gauge theory.
In fact, all the interpolating solutions have the same leading behavior for 
large values of $t$ (with the exception of the extremal points $\xi=1/6$ and $\xi=5/6$): 
using the asymptotic expansions given in section \ref{expansions}, one can write the leading contribution to the metric:
\beq
ds^2 \sim \frac{\textrm{const.}}{L^2}\, \frac{r^2}{\sqrt{\log r/r_0}}\,dx_m^2 + L^2\,\frac{d r^2}{r^2} \sqrt{\log r/r_0} + L^2 \sqrt{\log r/r_0}\, ds^2_{T^{1,1}}
\label{asym}
\eeq
where we have defined (to leading order) $r/r_0 \sim e^{t/3}$, and $d s^2_{T^{1,1}}$ is the standard metric on $T^{1,1}$:
\beq
ds^2_{T^{1,1}}=\frac{1}{6} \left[(e_1^2+e_2^2)+(\epsilon_1^2+\epsilon_2^2) \right] +\frac{1}{9}\tilde{\epsilon_3}^2
\eeq
The metric (\ref{asym}), apart from the logarithmic terms, is similar to the $AdS_5\times T^{1,1}$ metric with radius $L$ given by
\beq
L^2=\frac{9}{\sqrt{2}}\,\frac{M\alpha'}{2}e^{\phi_{UV}}
\eeq
where the last exponential represents the string coupling $g_s$, and where we have reintroduced the factor $M\alpha'/2$ that was considered 1 in previous calculations. 
The asymptotic form of the UV metric (\ref{asym}) depends on the flow parameter $\xi$ only by subleading corrections, suppressed by powers of $r$; also the RR and NS fields approach the asymptotic form of the KS solution. 
According to the standard philosophy of the AdS/CFT correspondence \cite{KW2}
we are lead to interpret our solution as a continuous family of vacua of
the $SU(N+M)\times SU(N)$ theory.

The corrections to the asymptotic KS metric and fluxes are interpreted as
the signal that expectation values of suitable operators are turned on.  
In the limit where $M\ll N$ we can be more precise. The gauge theory can be
then considered as approximatively conformal and we can use the field-operator
identification valid for the conformal theory associated 
with $AdS_5\times T^{1,1}$ \cite{KW1,ferrara}. Using the effective potential
(5.15) in \cite{PT} and expanding around the AdS vacuum ($P\rightarrow 0$) 
we can determine the masses of all the fields appearing in the PT ansatz 
in the conformal limit \cite{review}. We get the following values for the mass
squared (in units where the AdS radius $R=1$): $m^2=-4,-3,-3,0,0,12,21,32$,
which, using the usual relation $\Delta=2+\sqrt{4+m^2}$, corresponds to
a set of operators with dimension $\Delta=2,3,3,4,4,6,7,8$. The two marginal
operators are associated with $\phi$ and $h_1$ that determine the coupling 
constants of the two groups. Using the complete classifications of KK modes
on $T^{1,1}$ \cite{ferrara} we can tentatively identify the remaining fields as
operators in the following multiplets \cite{bigazzi,review}:
\bea
x,p \rightarrow {\mbox Tr} W^2 \bar W^2\,\,\, \Delta=6,8;&\qquad h_2,b\rightarrow
{\mbox Tr} W_1^2+W_2^2 , \,\,  {\mbox Tr}(A\bar A+B \bar B) W^2\,\,\, \Delta=3,7;&\nn\\
a\rightarrow {\mbox Tr} W_1^2-W_2^2\,\,\,\,\,\,\,\,\, \Delta=3;&\qquad g\rightarrow {\mbox Tr}A\bar A-B\bar B\,\,\,\,\,\,\,\,\,\,\qquad\qquad \Delta=2 \, .&
\label{op}
\eea  
where $W_i$ are the superfields in the vector multiplets for the two gauge groups, and $W$ is a combination of them
(sum or difference). $g$ in particular is associated
to the lowest component of the baryonic supercurrent \cite{KW2}.
KS solution has a non-zero vev for all of these operators
(for instance the vev for $h_2$ corresponds to a gaugino condensate
\cite{sonne}) except for the one corresponding
to $g$ which is $Z_2$ odd. Our family of deformations turns on a vev for this operator too, and it 
possibly modifies the ones that were already there in KS.
All the operators appearing in ~(\ref{op}) are $SU(2) \times SU(2)$ invariant thus
reflecting the symmetry of the baryonic branch. 
%When perturbing the KS solutions we change the existing VEVs and turn on 
%new ones for $Z_2$ odd operators.

The running of the couplings for the
two groups is determined by the $r$ dependence of the two functions
$e^\phi\sim 1/g_1^2+1/g_2^2$ and $h_1\sim 1/g_1^2-1/g_2^2$.
In the near conformal limit, the logarithmic running of the 
couplings found in \cite{KS} is only modified at subleading order
\bea
\frac{8\pi^2}{g_1^2}-\frac{8\pi^2}{g_2^2}&=&6 M \log (r/r_0)+O\left(\frac{1}{r^4}\right )\nn\\
\frac{8\pi^2}{g_1^2}+\frac{8\pi^2}{g_2^2}&=&{\mbox const}+O\left(\frac{1}{r^4}\right )
\label{coupling}
\eea
The leading order in $r$ matches the exact NSVZ formula for the 
$\beta$-function of supersymmetric gauge theories \cite{KS}. 
The extra terms, with
the AdS-inspired identification $r/r_s=\mu/\Lambda$, get the interpretation
of non-perturbative corrections. 

%All the previous reasonings are true along the flow for $1/6 < \xi <5/6$, 
%but when we reach the extremal MN solution the asymptotic behavior drastically changes, as we discussed in section \ref{expansions}, and the theory in the UV is even more exotic becoming effectively six dimensional.  

In the far infrared the behavior of the metric is always similar to the KS case: it reduces to the minkowskian space-time times an $S^3$, as expected for a deformed conifold. But now the radius of the three spheres depends on the flow parameter: using the small $t$ expansions given in section (\ref{expansions}) we obtain for the metric and fields:
\bea
ds^2 & \sim & e^{2A_0} d x_m^2+e^{\phi_0} M\alpha'\,\lambda \left( d\Omega^2_{S^3}+\frac{1}{4}\,dt^2\right) + \frac{e^{\phi_0}\lambda}{2}\,\frac{M\alpha'}{2}t^2(e_1^2+e_2^2) \label{uffa}\\
F_3 & \sim & -\frac{1}{2}\,\left( \frac{M\alpha'}{2} \right)\, \tilde{\epsilon}_1 \wedge \tilde{\epsilon}_2 \wedge \tilde{\epsilon}_3  \qquad \qquad H_3\sim 0
\label{strauffa}
\eea
where the $S^2$ sphere $e_1^2+e_2^2$ shrinks to zero and there are M units of RR flux through the $S^3$; $d\Omega^2_{S^3}$ is the metric of a round $S^3$ with unit radius:
\beq
d\Omega^2_{S^3} =\frac{1}{4} \left(  {\tilde{\epsilon}_1}^2 + {\tilde{\epsilon}_2}^2 + {\tilde{\epsilon}_3}^2\right)
\eeq

Fixing $\phi_{UV}$ and varying $\xi$, we can describe the entire baryonic 
branch of the $SU(N+M)\times SU(N)$ gauge theory. However,
as seen from Figure 1, fixing $\phi_{UV}$ forces the IR value $\phi_0$ to
(minus) infinity near the end of the flow. From eq.~(\ref{uffa}) we see that
the $S^3$ radius becomes small and the supergravity solution is strongly 
coupled. This means that we cannot trust the supergravity
description for large values of the baryonic VEV.
In particular, even though we found an interpolating solution between
KS and MN, we cannot claim that MN is at the endpoint of
the baryonic branch. What it is true is that we can connect KS and MN 
by varying simultaneously the baryonic VEV and a coupling constant, that is
by moving both in the space of theories and in the space of vacua.
 
Formulas~(\ref{uffa},\ref{strauffa}) 
are identical to equation (12) in \cite{herzog} (where our parameter $\lambda$ corresponds to their $b$); in that paper it was shown that $\lambda$ determines the tensions of k-strings, that are identified in supergravity with fundamental strings placed at $t=0$. One can therefore repeat the same supergravity calculation, finding for the tensions of k-strings, 
\beq T_k\sim \lambda \sin \psi \sqrt{1+(\lambda^2-1)\cos^2\psi} \label{ts}\eeq
where $\psi$ is the solution of the equation
\beq
\psi- \frac{\pi k}{M} =\frac{1-\lambda^2}{2}\sin (2\psi)\label{GGG}\eeq

The parameter $\lambda$ varies smoothly along the flow from its value at KS ($\lambda \sim 0.93266$) to MN ($\lambda=1$). The formula
\beq
\frac{T_q}{T_q'}=\frac{\sin \frac{\pi q}{M}}{\sin \frac{\pi q'}{M}}
\eeq  
is strictly valid only for MN, showing a non universality of the IR behavior.

\section{Conclusions}

In this paper we found a one-parameter family of
supersymmetric regular deformations of the Klebanov-Strassler solution. The existence of
these solutions
supports an older claim that in the last step of the KS cascade the gauge theory is
in the baryonic
branch. KS solution corresponds to a particular $Z_2$ symmetric point in this
baryonic branch,
while our family describes the whole moduli space. The solution becomes 
strongly coupled only for large values of the VEV.
This is the only known example where the moduli space of the gauge theory 
is described by a family of regular supergravity solutions. 
Moreover, by varying also the string coupling constant (that is by moving
also in the space of theories) we can smoothly connect the KS and MN
solutions. It would be interesting to understand what is the physical
meaning of this intepolating flow.

To find these solutions we made use of two tools: the interpolating ansatz for 
the metric and the
fluxes due to Papadopoulos and Tseytlin \cite{PT} and the supersymmetry conditions
obtained in \cite{GMPT}. 
Let us make a few comments about the ansatz, the method, and the solutions.

As always, the first order supersymmetry equations are easier to solve than the
equations of motion,
and that is what we are doing here. From the other side, as it is well known,
supersymmetry by itself
does not guarantee a solution, and Bianchi identities plus the equations of
motion for the fluxes have to be imposed. Typically it is difficult to find solutions of the
supersymmetry conditions that satisfy also the Bianchi identities.
In PT ansatz the fluxes are constructed in such a way that
 Bianchi identities are
automatically satisfied and
 we have explicitly checked that, for our solution, the supersymmetry conditions imply the equations of motions.
The use of SU(3) structures allows not simply to deal in a systematic way with the first order
equations, but breaks
everything into few basic representations and works representation by representation.
The PT ansatz together with the SU(3) structure (or spinorial) ansatz still have 13
undetermined
functions; our solution uses all of these. While the system of equations appears to
be heavily
overdetermined, there are many simplifications and eventually we were able to find
simple analytical
expressions for all the functions in terms of the solutions of two coupled
differential
equations. We solved these differential equations in power series for small and
large radius, having
thus the IR and UV asymptotics of the full solution.

Although supersymmetry did not forbid it, it was unclear that regular solutions with
SU(3) structure
besides KS and MN existed (as far as Polchinski-Strassler, we expect the exact
solution to have SU(2)
structure).  Realizing that GHK has SU(3) structure is a first step. Here we see
that the full
one-parameter family of solutions respects the SU(3) structure of the extrema. This
was a pleasant
surprise, which points toward a rich structure in the space of regular ${\cal N}=1$
supersymmetric solutions
with fluxes. As the SU(3) structure stays intact, throughout the family there is a
well-defined three form 
$\Omega$ without zeros such that $(d- W_5^{(\bar 3)}) \Omega = 0$. Moreover we can see
quite explicitly that in 
accordance with the general conditions for preserving ${\cal N}=1$ supersymmetry we can
shift $W_5$ away by scaling $\Omega$
and get a closed three-form (pure spinor),  and thus a family of generalized
Calabi--Yau manifolds.

We give the set of algebraic and differential equations governing the system. 
Although the full
analytical solution is still missing, the power expansion for small radius proves
the existence of a
one parameter family of regular solutions. 
All arbitrary constants in the solution
are fixed in terms
of one integration constant (plus another constant for the dilaton) which is the parameter along the flow. 
The flow parameter is allowed to take
values in an
interval that can be split into two regions related by a $Z_2$ symmetry. 
In the interpolating flow the KS solution
corresponds to the fixed point of this symmetry, while the MN solution is attained at
the extremal point
of the interval. The analysis of the asymptotic UV behavior shows that the whole
family behaves like
KS at large radius, except at the extremal MN point, where the UV behavior changes suddenly 
(for example, the dilaton blows up in
the UV).

From the gauge theory point of view, our family of solutions describes the baryonic branch of the confining vacua \cite{aharony,GHK}. 
Indeed, the large radius behavior of the supergravity fields
suggests that our solutions have an extra non-zero VEV with respect to KS for the
$Z_2$ odd operator $Tr(A \bar A - B \bar B)$. This behavior
confirms some expectations about the IR physics of KS 
\cite{aharony,GHK}: for example they are not
in the same IR universality class as the pure glue ${\cal N}=1$ gauge theory,
they are rather in a phase where there is confinement without mass gap.

As usual, with the supergravity solution at hand we can make qualitative and
quantitative predictions
on the strongly coupled regime of the dual gauge theory. Finding the supergravity
dual of the pure
glue ${\cal N}=1$ gauge theory remains a challenge, though, which means that the study of
supergravity
backgrounds with fluxes is far from being exhausted.

\vskip 1truecm

\noindent {\Large{\bf Acknowledgements}}

We would like to thank Nick Evans, Angel Paredes and Dimitrios Tsimpis for helpful discussions.
This work is supported in part by  INTAS grant 03-51-6346; RTN contracts MRTN-CT-2004-005104 and
MRTN-CT-2004-503369 and by a European Union Excellence Grant
MEXT-CT-2003-509661. MG was partially supported by European Commission Marie Curie Postdoctoral
Fellowship under contract number MEIF-CT-2003-501485. A. B. and A.Z. are
partially supported by INFN and MURST under contract 2001-025492, and by 
the European Commission TMR program HPRN-CT-2000-00131.

\vskip 1truecm

\noindent
{\Large{\bf Appendix A: Torsion and fluxes of PT ansatz in SU(3) representations}}
\renewcommand{\theequation}{A.\arabic{equation}}
\setcounter{equation}{0}\setcounter{section}{0}

\vskip 0.5truecm

We give the components of the torsion for the PT metric (\ref{PTmetric}) with the SU(3) structure given by (\ref{cs}) in the rotated basis:
\beq
\begin{array}{l@{\qquad}l@{\qquad}l}
G_1 = E_1 \, , &  G_2=\A E_2 + \B E_4 \, , & G_5=E_5  \\
G_3 = E_3 \, , &  G_4=\B E_2 - \A E_4  \, , & G_6=E_6  
\end{array}
\nn
\eeq

\bea
W_1 &=& \frac{1}{6} e^{-g-3 p-\frac{3 x}{2}}  \big(-\B+ a^2 \B-2 a \A e^g -\B e^{2
g}- 2 a\A e^{6 p+2x}  \nonumber \\
&-&  2\B e^{g+6p+2x} - 
2 a'e^{6p+2x} +  2 e^{g+6p+2x} (\B \A'  -  \A \B' )\big); \nn
\eea

\bea
W_2 &=& -\frac{2}{3}  e^{-g-3  p-\frac{3  x}{2} }  G_5 \wedge G_6    \
\big(- \B+ a^2  \B - 2 a  \A  e^g 
-  \B e^{2g}+    a  \A  e^{6  p+2  x} \nonumber \\
&+& \B  e^{g+6  p+2  x}+  e^{6  p+ 2  x} a'
- e^{g+6  p+2  x} (\B \A' - \A \B') \big) \nonumber \\
&+&  e^{-g+3p+\frac{x}{2}}  (G_2 \wedge G_3  -  G_1 \wedge G_4 )   
\big( a  \B +  \A  \B a' +   \B^2 e^g g' \big) \nonumber \\
&-& \frac{1}{3}  e^{-g-3  p- \frac{3  x}{2}}  G_1 \wedge G_2   
\big( \B-  a^2  \B+ 2 a  \A  e^g +  \B  e^{2  g}+ 2  a  \A  e^{6  p+2 x}
\nonumber \\
&-& \B  e^{g+6  p+2  x} + (3 \A^2 -1)   e^{6  p + 2  x} a'  + 
e^{g+6  p+2  x} (\B \A' - \A \B' + 3 \A \B g') \big) \nonumber \\
&+&  \frac{1}{3}  e^{-g-3  p- \frac{3  x}{2}}  G_3 \wedge G_4   
\big(- \B + a^2  \B - 2 a  \A  e^g -  \B  e^{2  g}+ 4  a  \A  e^{6  p+2 x}
\nonumber \\
&+& \B  e^{g+6  p+2  x} + (3 \A^2 + 1)   e^{6  p + 2  x} a'  + 
e^{g+6  p+2  x} (- \B \A' + \A \B' + 3 \A \B g') \big) \nn
\eea

\bea
W_3 &=&  -\frac{1}{4}  e^{-g- 3  p- \frac{3  x}{2}} ( G_1 \wedge G_3 
-  G_2 \wedge G_4 ) \wedge  G_6     
\big(-\B+ a^2  \B - 2  a  \A  e^g - \B  e^{2  g} \nonumber \\
&+&
 2  a  \A  e^{6  p+2  x}  +  
 2  \B  e^{g+6  p+2  x } - 2  e^{6  p+2 x}a'  +   2  e^{g+6  p+2  x} (\B \A' - \A
 \B')  \big)  \nonumber \\
&+&
\frac{\A}{2}e^{-g-3  p-\frac{3  x}{2}} ( G_1 \wedge G_2  -  G_3 \wedge G_4 )
\wedge G_5     \big(-1+{a^2}+e^{2  g}-2  \B  e^{6  p+2  x}a' +  2  \A 
 e^{g+6  p+2  x} g' \big) \nonumber \\
&+&
\frac{1}{4}  e^{-g-3  p-\frac{3  x}{2}}  G_2 \wedge G_3 \wedge G_5     
 \big(\B- {a^2}  \B-2  a  \A  e^g-3  \B  e^{2  g}+ 2  a 
\A  e^{6  p+2  x}  \nonumber \\
 &+&   2  \B  e^{g+6  p+2  x}+ 2 (\B^2 -\A^2)  e^{6  p+2 x} a'
 + 2    e^{g+6  p+2  x} (\B \A' - \A \B' - 2 \A \B g')\big) \nonumber \\
&+&
\frac{1}{4}  e^{-g-3  p-\frac{3  x}{2}}  G_1 \wedge G_4 \wedge G_5     
 \big(-3\B +3 {a^2}  \B - 2  a  \A  e^g +   \B  e^{2  g}+ 2  a 
\A  e^{6  p+2  x}  \nonumber \\
&+& 2  \B  e^{g+6  p+2  x} - 2 ( 1 + 2\B^2)  e^{6  p+2 x} a'
 + 2    e^{g+6  p+2  x} (\B \A' - \A \B' + 2 \A \B g') \big) \nn
\eea

\beq  
W_4 = \frac{1}{2}  e^{-g-3  p-\frac{3  x}{2}}  G_5  
 \big( - \A+ {a^2}  \A+2  a  \B  e^g - \A  e^{2 g}+  
2 e^{g+6  p+2  x}  x' \big) \nn
\eeq 

\beq  
 W_5^{(\bar 3)} =\frac{1}{4}  e^{-g+3p+\frac{x}{2}} (G_5-i G_6) 
\big(2  a  \B  - 2 \A  e^{g} - 6  e^{g} p' + e^{g} x' \big) \nn
\eeq

The components of the NS flux:

\bea
H^{(1)} &=&   \frac{1}{6}  e^{-g+3p-\frac{x}{2}} 
 \Big( (2  a  \A  e^g +\B 
(1-{a^2}+e^{2  g}))  \chi' +  (-2  a  \A  e^g + \B  
(1+{a^2}-e^{2  g})) h_1' \nonumber \\
&-&  2  (e^{g} h_2+ (-a  \B+ \A  e^g) h_2' ) \Big) \nn
\eea

\bea
H^{(3+\bar3)} &=&  \frac{1}{2}  e^{-g+3p-\frac{x}{2}}  G_5  
\Big((-2  a  \B  e^g +  \A  (1-{a^2}+e^{2  g})) \chi'+ 
 (\A+{a^2}  \A+2  a  \B  e^g- \A  e^{2  g}) h_1' \nn\\
&+&  
2  (a  \A+ \B  e^g) h_2' \Big) \nn
\eea

\bea
H^{(6+\bar6)} & =&
\frac{1}{4}  e^{-g+3p-\frac{x}{2}} \times \nn \\
\Big[&-&2  \A  ( G_1 \wedge G_2  - G_3 \wedge G_4 ) \wedge G_5   
((-1+{a^2}+e^{2  g}) \chi'
-  (1+{a^2}+e^{2  g})   h_1' -2  a h_2' ) 
\nonumber\\
&+&  G_2 \wedge G_3 \wedge G_5   (2  e^{g}  h_2
+ (2  a  \A  e^g +\B  (-1+{a^2}+3  e^{2  g})) \chi' \nonumber\\
&-&  (\B+{a^2}  \B+2  a  \A  e^g +3  \B  e^{2 g}) h_1'   - 
 2 ( a  \B  +  \A  e^g ) h_2' ) \nonumber \\
&+&  
 G_1 \wedge G_4 \wedge G_5   
(2  e^{g}  h_2
+  (2  a  \A  e^g - \B  (-3+ 3 {a^2}+  e^{2  g}))  \chi' \nonumber\\
&+& (3 \B+ 3 {a^2}  \B -  2  a  \A  e^g +  \B  e^{2 g}) h_1'   - 
 2 (-  3 a  \B  +  \A  e^g ) h_2' ) \nonumber \\
& -& ( G_1 \wedge G_3  - G_2 \wedge G_4 ) \wedge G_6      
 ((2  a  \A  e^g +\B  (1-{a^2}+e^{2  g} )) \chi' \nonumber\\
&+&  (-2  a  \A  e^g +\B  (1+{a^2}-e^{2  g})) h_1'+ 
 2  (e^{g} h_2 + (a  \B-\A  e^g)  h_2')) \Big] \nn
\eea

And finally the RR three-form flux components

\beq
F_3^{(1)} = -\frac{iP}{6}    e^{-g+3p-\frac{x}{2}}  (2  \A  (a-b)  e^g + 
  \B  (-1-{a^2}+2  a  b+e^{2  g}) -2  e^{g} b') \nn
\eeq

\beq
F_3^{(3+\bar3)} =
 \frac{P}{2}  e^{-g+3  p-\frac{x}{2}}    (2  (a-b)  \B  e^g+\A  
(1+{a^2}-2  a  b- e^{2  g}))   G_6 \nn
\eeq

\bea
F_3^{(6+\bar6)} &=&
\frac{P}{4}  e^{-g+3p-\frac{x}{2}} \Big[ 
2  \A   (1+{a^2}-2  a  b+e^{2  g})  
(G_1 \wedge G_2  - G_3 \wedge G_4) \wedge G_6  \nonumber\\
&+& 
G_1 \wedge G_4 \wedge G_6   (-2  \A  (a-b) e^g +  
\B  (3+3  {a^2}-6  a  b+e^{2  g}) -2  e^{g} b' ) \nonumber\\
&-&   G_1 \wedge G_3 \wedge G_5   (2  \A  (a-b) \
 e^g +  \B  (-1-{a^2}+2  a  b+e^{2  g}) +2  e^{g} b') \nonumber\\
& + &  G_2 \wedge G_4 \wedge G_5   (2  \A  (a-b) e^g 
+   \B  (-1-{a^2}+2  a  b+e^{2  g})+2  e^{g} b') \nonumber\\
&-&  G_2 \wedge G_3\wedge G_6  (2  \A  (a-b) e^g +
  \B  (1+{a^2}-2  a  b+3  e^{2  g})+2  e^{g}  b') \Big] \nn
\eea
%Here $P$ is arbitrary constant.

\vskip 0.5truecm
\noindent

{\Large{\bf Appendix B: Derivation of the susy equations}}
\renewcommand{\theequation}{B.\arabic{equation}}
\renewcommand{\thesubsection}{B.\arabic{subsection}}
\setcounter{equation}{0}\setcounter{section}{0}\setcounter{subsection}{0}

\vskip 0.5truecm

We derive here the susy equations for the PT ansatz imposing the conditions (\ref{cond1}) to (\ref{eq:intvec}).

\subsection{Conditions from $W_1$, $W_2$ and the singlets $H_3^{(1)}$, $F_3^{(1)}$}

Let's start with the conditions for the integrability of the complex structure (\ref{condcomp}). Setting to zero all the components of $W_1$ and $W_2$ given in the Appendix gives us five equations\footnote{As a check of our formalism, 
we compared our equations with the conditions
derived in \cite{PT}. The conditions derived by $W_1=W_2=0$ are equivalent to 
formulae (4.31) in \cite{PT} when
the identifications $e^{-g_{PT}}=2e^{-g}, t_{PT}=t/2$ and $(X,P)=(\A,\B)$ are made.}. After some algebra we reduce them to a set  
of only two independent equations: 
\bea 
2 \A a + \B e^{-g} (1-a^2+ e^{2g}) &=& 0 \, ,\label{complexcond1}  \\
a + a' \A + e^g g' \B &=& 0 \label{complexcond2} \, .
\eea 
Solving these for $\A$ and $\B$ and imposing the constraint $\A^2+\B^2=1$ gives an equation for $g'$
\beq
g'= e^{-2\,g} \left[ a\,S + \left( C - a \right) \,a' \right] \, ,
\label{derg}
\eeq
where we have defined the following useful quantities:
\bea
S & \equiv & \frac{{\sqrt{a^4 + 2\,a^2\,\left( -1 + e^{2\,g} \right)  + {\left( 1 + e^{2\,g} \right) }^2}}}{2\,a}\label{sinh} \, ,\\
C & \equiv & \frac{1 + a^2 + e^{2\,g}}{2\,a}\label{cosh} \, .
\eea
From these definitions it follows that $C^2-S^2=1$.
Differentiating (\ref{cosh}), (\ref{sinh}) and using only the equation (\ref{derg}), one can show that $C$ and $S$ satisfy the remarkable conditions: $C'=S$ and $S'=C$, allowing us to integrate them
\bea
C & = & -\left( k_1 \cosh t + k_2 \sinh t \right) \label{ch} \, , \\
S & = & -\left( k_1 \sinh t + k_2 \sinh t \right) \label{sh} \, .
\eea
These are two algebraic relations that determine $e^{2g}$ in terms of $a$ and $t$. From $C^2-S^2=1$ we find the constraint: $k_1^2-k_2^2=1$.
% this is equivalent to the statement $C^2-S^2=1$ (obviously we expected only one independent integration constant, since we started with a single first order equation (\ref{derg})). 
Parameterizing $k_1$ and $k_2$ as $k_1=\cosh t_0$, $k_2=\sinh t_0$, equations (\ref{ch}), (\ref{sh}) become: $C=-\cosh (t+t_0)$, $S=-\sinh (t+t_0)$. Therefore, up to a redefinition of $t$, we can always fix the integration parameters to $k_1=1$, $k_2=0$, corresponding to
\beq
C=-\cosh t \, , \qquad S=-\sinh t \, .
\eeq 
Notice that the expressions (\ref{MN1}), (\ref{KS1}), commonly found in the literature, are written with this choice of $t$ and they have $0 \leq t<+\infty$. With this fixing the expression of $g$ as a function  of $a$ (\ref{ch}) becomes: 
\beq
e^{2g}=-1 - a^2 - 2\,a\,\cosh t \label{gint} \, .
\eeq
From conditions (\ref{complexcond1}), (\ref{complexcond2}) and using (\ref{derg}), we can derive the following expressions for $\A$, $\B$:
\beq
\A=\frac{C-a}{S}  \qquad \mbox{and}  \qquad \B=-\frac{e^{g}}{S} \, , 
\label{AB}
\eeq
which  will be used to eliminate $\A$ and $\B$ from following formulas.  

Let's turn now to the singlets conditions on $H^{(1)}$ and  $F_3^{(1)}$(\ref{cond1}); they give differential equations for $h_1'$ and $b'$,  
respectively:
\bea
h_1' & = & -h_2 S -h_2' C \, , \label{derh1} \\
b' & = & \frac{1-b\,C}{S} \, . \label{derb}
\eea
Note that using ($\ref{AB}$), $\chi'$ drops out of the first equation. They can both be integrated to give:
\bea
h_1 & = & -h_2 C + \tilde Q \, ,\\
b & = & \frac{t + c_b}{S} \, ,
\eea  
where $c_h$ and $c_b$ are integration constants; notice that imposing regularity conditions in $t=0$ forces us to put $c_b=0$. 
%Instead $Q$ 
%does not appear in our equations, since only $h_1'$ appears in the PT ansatz for the field strength $H$. 
%{(\bf check! see also below...does Q have some physical meaning? or can always be gauged away?)}
 With the choice $k_1=1$, $k_2=0$, $c_b=0$, the previous expressions become:
\bea
h_1 & = & h_2 \cosh t+\tilde Q \, ,  \label{h1int} \\
b & = & -\frac{t}{\sinh t} \, .\label{bint}
\eea 
Notice in particular that $b$ does not vary along the flow.

Summarizing, we managed to integrate the equations for $g$, $b$, $h_1$, so that they can  always be thought as functions of $a$ and $h_2$. In fact we will use their integrated expressions only in the last steps (\ref{coupled}) in order to avoid the introduction of integration parameters from the beginning, and so, to simplify following formulas, we will use only their differential expressions.  

\subsection{Conditions from the 6 sector}
Analyzing the conditions (\ref{eq:6int}) (note for example that the third one can be derived from the first two eliminating $W_3$), it is easy to show from their explicit expressions in the Appendix that they give four independent equations which  can be rearranged in two algebraic expressions:
\bea
\sin w & = & 2\, e^{x-g-\phi} \, \frac{a\,C - 1}{b\,C - 1} \, , \label{equsin}\\
\cos w & = & 2\, e^{-\phi} \, h_2 \, \frac{S}{b\,C -1} \, ,\label{equcos}
\eea
and two differential equations for $a'$ and $\chi'$:
\bea
a' & = & - \frac{\left( a\,C -1 \right)}{S} \,e^{g - 2x-6p} +
\frac{a\,\left( a - b \right) \,S}{b\,C-1} \, , \label{equa}\\
\chi' & = & - a\,S\,\frac{\left( b - 2\,C + b\,C^2 \right) \,h_2 + 
        \left( b\,C-1 \right) \,S\,h_2' }{\left( a\,C-1 \right) \,
      \left( b\,C-1 \right) } \, .
\label{chi1}      
\eea
Notice that since the equations (\ref{eq:6int}) are homogeneous in $\alpha$ and $\beta$, the algebraic expressions depend only upon $w$ defined as in 
(\ref{par}): $\beta=i\,\alpha \tan(w/2)$. 

\subsection{Conditions from the $\bar{3}$ sector}

The strategy we used to analyze  equations (\ref{flux-vector}) and (\ref{eq:intvec}) was to write first the ratio of every expression to the first equation of (\ref{flux-vector}) (the one for $F_3^{(\bar{3})}$), in order to simplify the dependence on $\bar{\partial}\alpha$.

The ratio between the equations for $F_5^{(\bar{3})}$ and $F_3^{(\bar{3})}$ translates into the condition for the function $K$ in front of the PT ansatz for $F_5$:
\beq
K=h_2(C-b) \qquad \Rightarrow \qquad K=-h_1-b h_2+\tilde Q \, ,
\eeq
where in the second equality we have used the integrated expression for $h_1$ in order to show that this susy equation coincides with the only non trivial Bianchi identity: $dF_5=H\wedge F$, that is $K'=-(h_1+b h_2)'$. $\tilde Q$ is then
identified with the quantity $Q$ defined in the PT ansatz.

Note that the expression $h_2(C-b)$ for the five form $F_5$ is always vanishing at $t=0$ for all the regular interpolating solutions we discussed in Section (\ref{section4}): inserting the small $t$ expansion (\ref{serieh2}) we find $K\propto t^3$. In the language of gauge theory this should mean that the solutions found in this paper all correspond to $p=0$ where $N=nM+p$ and $0 \leq p <M$, with $N$ and $M$ the number of regular and fractional branes respectively. This is the only value of $p$ where a baryonic branch exists. 
%We could have expected this fact since we have discarded singular solutions in $t=0$ and we know from \cite{KS} that trying to introduce a non zero $F_5$ flux at $t=0$ translates (for the KS solution) in a singular horizon in $t=0$ (see also the discussion in paragraph 5.2 of \cite{KS}).
   
From the ratio of the equations $H_3^{(\bar{3})}$ and $F_3^{(\bar{3})}$ we derive
\beq
\chi'=-\frac{\left( a\,C-1 \right) \,\left( b\,h_2\,S + 
        \left( b\,C-1 \right) \,h_2' \right) }{a\,\left( b\,C-1 \right) }  \, . 
\label{chi2}        
\eeq

Equations (\ref{chi1}) and (\ref{chi2}) can be rearranged in 
\bea
h_2' & = & \frac{ 2\,a^2\,C - b\,\left( -1 + a^2 + 2\,a\,C \right)  }
  { b\,C-1 }\,e^{-2\,g}\,h_2\,S \, ,\label{equh2}\\
\chi' & = & \frac{2\,a \left( b-C \right) \left( a\,C-1 \right)}{b\,C-1}\,e^{-2\,g}\,h_2\,S \, ,
\label{chi3}
\eea
and it is easy to show that equation (\ref{chi3}) is equal to the equation of motion for $\chi$\footnote{This is the equation (5.21) in \cite{PT}.}.

The ratios of the equations for $W_4$, $W_5^{(\bar 3)}$, $\bar{\partial}\phi$, $\bar{\partial} A$ with the equation for $F_3^{\bar{3}}$ give respectively the following differential equations for $x'$, $p'$, $\phi'$, $A'$: 
\bea
x' & = & a\,S\,e^{-g - 6\,p - 2\,x} + \frac{b - C}{b\,C-1}\,h_2^2\,S\,e^{-2\,x} \, ,\label{equx}\\
p' & = & -\frac{e^{-2\,g}}{12\,S\,\left( b\,C-1 \right)} \Big[ e^{-2\,x-6\,p} \Big( 
  4\,\left( b - C \right) \,\left( -1 - a^2 + 2\,a\,C \right) \,e^{6\,p}\,{h_2}^2\,S^2  \nn \\ & &
    -2\,a\,\left( b\,C-1 \right) \,e^g\,S^2 \Big)- \Big( 4\,a + 2\,b + 2\,a^2\,b - 2\,C - 2\,a^2\,C   \, ,
  \nn \label{equp}\\ & &  
   - 4\,a\,b\,C + 4\,a\,S^2 + 4\,b\,S^2 + 2\,a^2\,b\,S^2 + 2\,a^2\,C\,S^2 - 8\,a\,b\,C\,S^2 \Big) \Big]\\
\phi' & = & \frac{\left( C-b \right) \,{\left( a\,C-1 \right) }^2}{\left( b\,C-1 \right) \,S}\,e^{-2\,g} \, ,\label{equphi}\\
A' & = & \frac{ b - C - b^2\,C + b\,C^2 }{8\,S}\,e^{-2\,x + 2\,\phi } \, .\label{equA}
\eea

There are still two conditions to be imposed:
\beq
A=\log \left( |\alpha|^2+|\beta|^2 \right) 
\label{alphabeta}
\eeq
which determines the values of $\alpha$ and $\beta$ once $A$ and $w$ are known, and finally the equation (\ref{flux-vector}) for $F_3^{(\bar{3})}$. If we use the derivative of (\ref{alphabeta}) to express $\alpha'$ in function of $w'$, we get from $F_3^{(\bar{3})}$ the following differential equation for $w'$:
\beq
w' = \frac{\left( b - C \right) \,\left( a\,C-1 \right) }{ b\,C-1 }\,e^{-g - x}\,h_2 \, .\label{equw}
\eeq
 
\subsection{Check of consistency and other useful relations}

If we treat f $g$, $b$, $h_1$, $\A$, $\B$ as functions of $a$ and $h_2$, we are left with the following unknowns: $a$, $h_2$, $\phi$, $x$, $p$ and $w$. Note in fact that $A$, $\chi'$, $\alpha$ and $\beta$ are determined from the equations (\ref{equA}), (\ref{chi3}), (\ref{alphabeta}) once the other quantities are known and do not enter in other equations. So we are left with 6 unknowns and 8 equations: the two algebraic (\ref{equsin}), (\ref{equcos}), and the six differential (\ref{equa}), (\ref{equh2}), (\ref{equx}), (\ref{equp}), (\ref{equphi}), (\ref{equw}).

The system may seem overdetermined, but it is not so: it is straightforward to show that the equations for $x'$ (\ref{equx}) and for $w'$ (\ref{equw}) may be obtained respectively by differentiating the two algebraic equations (\ref{equsin}), (\ref{equcos}) and using the other equations. We may therefore discard these two equations and write a set of independent equations for $a$, $h_2$, $\phi$, $x$, $v\equiv e^{2x+6p}$ and $w$ as:
\bea
a' & = & - \frac{\left( a\,C -1 \right)\,e^{g}}{S} \,\frac{1}{v} +
\frac{a\,\left( a - b \right) \,S}{b\,C-1}\, , \\
v' & = & 3\,a\,e^{-g}\, S + v \, \frac{e^{-2\,g}}{\left( b\,C-1 \right)\,S} \, \Big[ b\,\left( -1 + 2\,a\,C + 2\,C^2 + a^2\,C^2 - 4\,a\,C^3 \right) \nn \\ & &
 + \, C\,\left( -1 + 2\,a\,C + a^2\,\left( -2 + C^2 \right)  \right) \Big] \, ,\label{equv}\\
\phi' & = & \frac{\left( C-b \right) \,{\left( a\,C-1 \right) }^2}{\left( b\,C-1 \right) \,S}\,e^{-2\,g} \, , \\
h_2' & = & \frac{ 2\,a^2\,C - b\,\left( -1 + a^2 + 2\,a\,C \right)  }
  { b\,C-1 }\,e^{-2\,g}\,h_2\,S \, ,\label{equh22}\\ 
\sin w & = & 2\, e^{x-g-\phi} \, \frac{a\,C - 1}{b\,C - 1} \, ,\\
\cos w & = & 2\, e^{-\phi} \, h_2 \, \frac{S}{b\,C -1}   \, ,
\eea
where the equation for $v$ follows directly from (\ref{equx}),  (\ref{equp}).
Note that the first two equations are a coupled first order system that determines $a$ and $v$; the next two determine $\phi$ and $h_2$, and the algebraic ones determine $w$ and $x$. Alternatively it is easy to show that one could discard the differential equation for $h_2$ (\ref{equh22}) and keep instead that for $x'$ (\ref{equx}).
All these susy equations are satisfied for the KS, MN and GHK solutions.

We can perform a further integration for the dilaton equation: note that from (\ref{equphi}), (\ref{equw}) and the algebraic relations it follows that:
\beq
\phi'=-w'\tan w 
\eeq
which  can be integrated to
\beq
e^{\phi}=\frac{\cos w}{\eta} \, ,
\label{phiw}
\eeq 
with $\eta$ an integration parameter; notice that this cannot be done for the extremal MN case, since for MN $w=\pi/2$ and consequently $\eta\rightarrow 0$. 
The relation (\ref{phiw}) determines $w$ in terms of $\phi$; we can rewrite the previous set of equations for the variables $a$, $v$, $\phi$, $h_2$, $x$ discarding the dilaton equation:
\bea
a' & = & - \frac{\left( a\,C -1 \right)\,e^{g}}{S} \,\frac{1}{v} +
\frac{a\,\left( a - b \right) \,S}{b\,C-1} \, ,\\
v' & = & 3\,a\,e^{-g}\, S + v \, \frac{e^{-2\,g}}{\left( b\,C-1 \right)\,S} \, \Big[ b\,\left( -1 + 2\,a\,C + 2\,C^2 + a^2\,C^2 - 4\,a\,C^3 \right) \nn \\ & &
 + \, C\,\left( -1 + 2\,a\,C + a^2\,\left( -2 + C^2 \right)  \right) \Big] \, ,\\
h_2' & = & \frac{ 2\,a^2\,C - b\,\left( -1 + a^2 + 2\,a\,C \right)  }
  { b\,C-1 }\,e^{-2\,g}\,h_2\,S \, , \\ 
e^{2 \phi} & = & \frac{2}{\eta} \, \frac{S}{b\,C-1} \, h_2 \, ,\\
e^{2x} & = & \left(\frac{e^g\,S}{a\,C-1}\right)^2 \left(\frac{1}{e^{2\phi}\eta^2}-1 \right) \,  h_2^2 \, ,
\eea
where we have eliminated $w$ from the algebraic expressions (\ref{equsin}), (\ref{equcos}) through (\ref{phiw}) and solved with respect 
to $\phi$ and $x$.
Notice that the relation between $\eta$ and the other integration constants is:
\beq
\eta= d\,e^{-2\,\phi_{UV}}=e^{-\phi_0}{\sqrt{1 - \frac{9\,{\left( 1 - 2\,\xi  \right) }^2\,{\lambda }^2}{4}}} \, .
\eeq
In conclusion let's say few words as one can verify that equations of motion are automatically satisfied by our susy equations. We do not report
 here the calculations since they are long but purely algebraic. We already said that Bianchi identities are satisfied by the PT ansatz. Then one has to write all the equations of motion for the metric, fluxes and dilaton (the equation for the axion is trivially satisfied by the PT ansatz with $C_0=0$) in string frame. This is a system of second order differential equations much more complicated than the susy equations for the functions $A$, $x$, $p$, $g$, $a$, $h_1$, $h_2$, $\chi$, $b$, $\phi$; obviously they do not depend on the functions from supersymmetry and complex structure $\alpha$, $\beta$, $w$, $\A$, $\B$. Replacing every second and first derivative of the functions in the equations of motion with the derivatives of the susy differential equations (\ref{equA}), (\ref{equx}), (\ref{equp}), (\ref{derg}), (\ref{equa}), (\ref{derh1}), (\ref{equh2}), (\ref{chi3}), (\ref{derb}), (\ref{equphi}) or with the susy differential equations themselves, one finds only algebraic relations that, after simplification, can always be reduced to the identity using the susy algebraic equation that can be deduced eliminating $w$ from (\ref{equsin}), (\ref{equcos}) through $\sin^2w+\cos^2w=1$.


\begin{thebibliography}{99}

\bibitem{GMPT}
M.~Gra\~na, R.~Minasian, M.~Petrini and A.~Tomasiello,
``Supersymmetric backgrounds from generalized Calabi-Yau manifolds,''
JHEP {\bf 0408}, 046 (2004)
[arXiv:hep-th/0406137].
 

\bibitem{KS}
I.~R.~Klebanov and M.~J.~Strassler,
``Supergravity and a confining gauge theory: Duality cascades and
chiSB-resolution of naked singularities,''
JHEP {\bf 0008}, 052 (2000)
[arXiv:hep-th/0007191].

\bibitem{MN}
J.~M.~Maldacena and C.~Nu\~nez,
``Towards the large N limit of pure N = 1 super Yang Mills,''
Phys.\ Rev.\ Lett.\  {\bf 86}, 588 (2001)
[arXiv:hep-th/0008001].

\bibitem{aharony}
O.~Aharony,
``A note on the holographic interpretation of string theory backgrounds  with
varying flux,''
JHEP {\bf 0103} (2001) 012
[arXiv:hep-th/0101013].


\bibitem{GHK}
S.~S.~Gubser, C.~P.~Herzog and I.~R.~Klebanov,
``Symmetry breaking and axionic strings in the warped deformed conifold,''
JHEP {\bf 0409}, 036 (2004)
[arXiv:hep-th/0405282].

\bibitem{klebstring} 
S.~S.~Gubser, C.~P.~Herzog and I.~R.~Klebanov,
``Variations on the warped deformed conifold,'' C. R. Physique 5 (2004)
[arXiv:hep-th/0409186].

\bibitem{FG}
A.~R.~Frey and M.~Gra\~na,
``Type IIB solutions with interpolating supersymmetries,''
Phys.\ Rev.\ D {\bf 68}, 106002 (2003)
[arXiv:hep-th/0307142].


\bibitem{Frey}
A.~R.~Frey,
``Notes on SU(3) structures in type IIB supergravity,''
JHEP {\bf 0406}, 027 (2004)
[arXiv:hep-th/0404107].



\bibitem{DAllagata}
G.~Dall'Agata,
``On supersymmetric solutions of type IIB supergravity with general fluxes,''
Nucl.\ Phys.\ B {\bf 695}, 243 (2004)
[arXiv:hep-th/0403220].


\bibitem{PT}
G.~Papadopoulos and A.~A.~Tseytlin,
``Complex geometry of conifolds and 5-brane wrapped on 2-sphere,''
Class.\ Quant.\ Grav.\  {\bf 18}, 1333 (2001)
[arXiv:hep-th/0012034].


\bibitem{ds}
M.~R.~Douglas and S.~H.~Shenker,
``Dynamics of SU(N) supersymmetric gauge theory,''
Nucl.\ Phys.\ B {\bf 447} (1995) 271
[arXiv:hep-th/9503163].
\bibitem{HSZ} 
A.~Hanany, M.~J.~Strassler and A.~Zaffaroni,
``Confinement and strings in M{QCD},''
Nucl.\ Phys.\ B {\bf 513} (1998) 87
[arXiv:hep-th/9707244].

\bibitem{herzog}
C.~P.~Herzog and I.~R.~Klebanov,
``On string tensions in supersymmetric SU(M) gauge theory,''
Phys.\ Lett.\ B {\bf 526}, 388 (2002)
[arXiv:hep-th/0111078].

\bibitem{JPP}
C.~V.~Johnson, A.~W.~Peet and J.~Polchinski,
``Gauge theory and the excision of repulson singularities,''
Phys.\ Rev.\ D {\bf 61}, 086001 (2000)
[arXiv:hep-th/9911161];

A.~Buchel, A.~W.~Peet and J.~Polchinski,
``Gauge dual and noncommutative extension of an N = 2 supergravity
solution,''
Phys.\ Rev.\ D {\bf 63}, 044009 (2001)
[arXiv:hep-th/0008076];

N.~J.~Evans, C.~V.~Johnson and M.~Petrini,
``The enhancon and N = 2 gauge theory/gravity RG flows,''
JHEP {\bf 0010}, 022 (2000)
[arXiv:hep-th/0008081];

D.~Z.~Freedman, S.~S.~Gubser, K.~Pilch and N.~P.~Warner,
``Continuous distributions of D3-branes and gauged supergravity,''
JHEP {\bf 0007}, 038 (2000)
[arXiv:hep-th/9906194].


\bibitem{Salomon}
S. Salamon, {\it Riemannian Geometry and Holonomy Groups}, Vol.~201 of
{\it Pitman Research Notes in Mathematics}, Longman, Harlow, 1989.

\bibitem{joyce}
D.\ Joyce, ``Compact Manifolds with Special Holonomy", Oxford University Press,
Oxford, 2000.

\bibitem{London}
J.~P.~Gauntlett, D.~Martelli and D.~Waldram,
``Superstrings with intrinsic torsion,''
Phys.\ Rev.\ D {\bf 69}, 086002 (2004)
[arXiv:hep-th/0302158].




\bibitem{GP}
M.~Gra\~na and J.~Polchinski,
``Supersymmetric three-form flux perturbations on AdS(5),''
Phys.\ Rev.\ D {\bf 63}, 026001 (2001)
[arXiv:hep-th/0009211].

\bibitem{Gubser}
S.~S.~Gubser,
``Supersymmetry and F-theory realization of the deformed conifold with
three-form flux,''
arXiv:hep-th/0010010.


\bibitem{Strominger}
A.~Strominger,
``Superstrings With Torsion,''
Nucl.\ Phys.\ B {\bf 274}, 253 (1986);
C.~M.~Hull,
``Superstring Compactifications With Torsion And Space-Time Supersymmetry,''
Print-86-0251 (CAMBRIDGE)

\bibitem{Hitchin}
N.~Hitchin, ``The geometry of three-forms in six and seven dimensions,'' 
arXiv:math.DG/0010054.
 


\bibitem{MT}
R.~Minasian and D.~Tsimpis,
``On the geometry of non-trivially embedded branes,''
Nucl.\ Phys.\ B {\bf 572}, 499 (2000)
[arXiv:hep-th/9911042].

\bibitem{CIV}
F.~Cachazo, K.~A.~Intriligator and C.~Vafa,
``A large N duality via a geometric transition,''
Nucl.\ Phys.\ B {\bf 603}, 3 (2001)
[arXiv:hep-th/0103067].

\bibitem{apreda}
R.~Apreda,
``Non supersymmetric regular solutions from wrapped and fractional branes,''
arXiv:hep-th/0301118.

\bibitem{KW2}
I.~R.~Klebanov and E.~Witten,
``AdS/CFT correspondence and symmetry breaking,''
Nucl.\ Phys.\ B {\bf 556}, 89 (1999)
[arXiv:hep-th/9905104].

\bibitem{KW1}
I.~R.~Klebanov and E.~Witten,
``Superconformal field theory on threebranes at a Calabi-Yau  singularity,''
Nucl.\ Phys.\ B {\bf 536}, 199 (1998)
[arXiv:hep-th/9807080].

\bibitem{ferrara}
A.~Ceresole, G.~Dall'Agata, R.~D'Auria and S.~Ferrara,
``Spectrum of type IIB supergravity on AdS(5) x T(11): Predictions on N  = 1
%SCFT's,''
Phys.\ Rev.\ D {\bf 61}, 066001 (2000)
[arXiv:hep-th/9905226].

\bibitem{bigazzi}
F.~Bigazzi, L.~Girardello and A.~Zaffaroni,
``A note on regular type 0 solutions and confining gauge theories,''
Nucl.\ Phys.\ B {\bf 598} (2001) 530
[arXiv:hep-th/0011041].
\bibitem{review}
F.~Bigazzi, A.~L.~Cotrone, M.~Petrini and A.~Zaffaroni,
``Supergravity duals of supersymmetric four dimensional gauge theories,''
Riv.\ Nuovo Cim.\  {\bf 25N12}, 1 (2002)
[arXiv:hep-th/0303191].

\bibitem{sonne}
A.~Loewy and J.~Sonnenschein,
``On the holographic duals of N = 1 gauge dynamics,''
JHEP {\bf 0108}, 007 (2001)
[arXiv:hep-th/0103163].

\end{thebibliography}
\end{document}